\documentclass[lettersize,journal]{IEEEtran}
\usepackage{nopageno}
\ifCLASSINFOpdf
\else
\fi

\usepackage{cite}
\usepackage{amsmath,amssymb,amsfonts}
\usepackage{mathtools} 
\usepackage{algorithmic}
\usepackage[ruled,vlined,linesnumbered]{algorithm2e}
\usepackage{graphicx}
\usepackage{textcomp}
\usepackage{authblk}
\usepackage{blindtext}
\usepackage{enumitem}
\usepackage{subcaption}
\usepackage{booktabs}
\usepackage{caption} 
\usepackage{float}
\usepackage{xcolor}

\floatstyle{plaintop}
\restylefloat{table}
 
\usepackage[font=footnotesize]{caption}
\DeclareMathOperator*{\argmax}{arg\,max}

\begin{document}
\title{Time Efficient Joint UAV-BS Deployment and User Association based on Machine Learning}
\author{Bo~Ma,
Zitian~Zhang,
Jiliang~Zhang,~\IEEEmembership{Senior Member,~IEEE,}
and~Jie~Zhang,~\IEEEmembership{Senior Member,~IEEE}

\thanks{Manuscript received ; revised .}
\thanks{Bo Ma is with the School of Information and Electronic Engineering, Zhejiang Gongshang University, Hangzhou, 310018, China (e-mail: mabo@mail.zjgsu.edu.cn).}

\thanks{Jiliang Zhang, and Jie Zhang are with the Department of Electronic and Electrical Engineering, The University of Sheffield, Sheffield S1 3JD, U.K. (e-mail: \{jiliang.zhang, jie.zhang\}@sheffield.ac.uk).}

\thanks{Zitian Zhang, and Jie Zhang are with Ranplan Wireless Network Design Ltd, Cambridge, CB23 3UY, U.K. (e-mail: \{zitian.zhang, jie.zhang\}@ranplanwireless.com). (Zitian Zhang and Jie Zhang are the co-corresponding authors.)}
       
\thanks{This work was supported in part by}


}

\markboth{Journal of \LaTeX\ Class Files,~Vol.~14, No.~8, August~2021}%
{Shell \MakeLowercase{\textit{et al.}}: A Sample Article Using IEEEtran.cls for IEEE Journals}


\maketitle

\vspace{-1.5cm}

\begin{abstract}
This paper proposes a time-efficient mechanism to decrease the on-line computing time of solving the joint unmanned aerial vehicle base station (UAV-BS) deployment and user/sensor association (UDUA) problem aiming at maximizing the downlink sum transmission throughput. The joint UDUA problem is decoupled into two sub-problems: one is the user association sub-problem, which gets the optimal matching strategy between aerial and ground nodes for certain UAV-BS positions; and the other is the UAV-BS deployment sub-problem trying to find the best position combination of the UAV-BSs that make the solution of the first sub-problem optimal among all the possible position combinations of the UAV-BSs. In the proposed mechanism, we transform the user association sub-problem into an equivalent bipartite matching problem and solve it using the Kuhn-Munkres algorithm. For the UAV-BS deployment sub-problem, we theoretically prove that adopting the best UAV-BS deployment strategy of a previous user distribution for each new user distribution will introduce little performance decline compared with the new user distribution's ground true best strategy if the two user distributions are similar enough. Based on our mathematical analyses, the similarity level between user distributions is well defined and becomes the key to solve the second sub-problem. Numerical results indicate that the proposed UDUA mechanism can achieve near-optimal system performance in terms of average downlink sum transmission throughput and failure rate with enormously reduced computing time compared with benchmark approaches.

\end{abstract}

%
\IEEEpeerreviewmaketitle

\section{Introduction}
\IEEEPARstart{W}{ith} the fast development of information industry and Internet for everything, unprecedented demands of high-quality wireless services are imposing enormous challenges to mobile networks. Unmanned aerial vehicles (UAVs) carrying aerial base stations have been widely utilized to enhance the service provisioning of the existing terrestrial communication infrastructure \cite{saad20216g}, especially for emerging scenarios such as 
data exchange in Internet of Things (IoT) systems and fast-response mobile network assistance \cite{zeng2021uav,Ht2021iot,Li2022iot}. These scenarios are likely of high sensor/device density and the deployment of multiple UAV base stations (UAV-BSs) hovering in fixed positions during the transmission period can provide stable and continuous wireless services to ground devices.

Despite the potential benefits of UAV-BSs in establishing flexible and on-demand wireless connections via likely line-of-sight (LoS) links to ground users \cite{Bor-Yaliniz2016}, the deployment of UAV-BSs is still facing some key challenges. On the one hand, the channel conditions between UAV-BSs and ground users are highly influenced by their relative locations, indicating that the UAV-BS deployment and user association (UDUA) strategies need to be jointly designed. On the other hand, to provide on-demand wireless services, the UDUA strategies must be calculated in a time-efficient way.

The UDUA problem has been widely investigated in recent years to improve system performance in terms of UAV-BS coverage, energy efficiency, and uplink/downlink transmission rate\cite{Mozaffari2017}-\cite{Abeywickrama2020}. However, the existing UDUA approaches handle each UDUA problem individually and rely on complex algorithms to obtain the optimal or sub-optimal solution for each specific UDUA problem. The high computational complexity renders it impossible for these algorithms to respond swiftly to service demand as typically expected for UAV-BSs.

In this paper, we propose to maximize the downlink sum transmission rate for the ground users served by multiple UAV-BSs while guaranteeing the quality of service (QoS) for each ground user. More specifically, we develop a centralized mechanism to solve the UDUA problem before the UAVs are dispatched. In order to reduce the on-demand response time, the experiences are accumulated from previously solved UDUA problems to acquire the proper UAV-BS deployment strategy for a new UDUA problem. After the UAV-BS positions are determined, the optimal associations between the ground users and the UAV-BSs are then obtained by solving an equivalent bipartite matching problem. The main contributions of this paper are summarized as follows:
\IEEEpubidadjcol
\begin{itemize}
\item We maximize the downlink sum transmission rate of the ground users distributed in a certain region by jointly optimizing the UAV-BS positions and the association between the UAV-BSs and the ground users. By dividing the considered region into small grids and modeling the UAV-BS positions as discrete variables, we formulate the joint UDUA problem into an integer non-linear programming (INLP) problem subject to the QoS requirement of each ground user.
\item Since the user association can be decided after the positions of UAV-BSs have been determined, we decouple the joint UDUA problem into two sub-problems. One is the user association sub-problem looking for the optimal matching strategy between the UAV-BSs and the ground users for every possible combination of UAV-BS positions. The other is the UAV-BS deployment sub-problem searching the best combination of UAV-BS positions that returns the maximum downlink sum rate among all the possible combinations of UAV-BS positions when they are combined with their optimal user association strategies.
\item We propose a centralized UDUA mechanism to solve the above two sub-problems. In particular, we transform the user association sub-problem into an equivalent bipartite matching problem and solve it using the Kuhn-Munkres algorithm. For the UAV-BS deployment sub-problem, we theoretically prove that adopting the best UAV-BS deployment strategy of a previous user distribution for each new user distribution will introduce little performance decline compared with the new user distribution's ground true best strategy if the two user distributions are similar enough. Based on our mathematical analyses, the similarity level between user distributions is well defined and a k-nearest neighbor (KNN) based algorithm is presented to solve the second sub-problem.
\item We evaluate the proposed mechanism through extensive experiments. Numerical results indicate that the proposed UDUA mechanism can achieve near-optimal system performance in terms of average downlink sum transmission rate and failure rate with enormously reduced computing time compared with existing UDUA approaches.
\end{itemize}

The rest of this paper is organized as follows: In Section II, related works are reviewed. Section III provides the system model and the optimization problem formulation. In Section IV, the proposed UDUA mechanism is elaborately introduced. In Section V, we evaluate the proposed mechanism's performance. Finally, Section VI concludes this paper.

\vspace{-0.0cm}
\section{Related Works}
According to the approach used to solve the UDUA problem, existing UDUA approaches can roughly be divided into two categories, i.e., the model-driven UDUA approaches and the machine learning based approaches.

In the first category, the UDUA problem is solved using the convex optimization tools or the modern optimization algorithms. Focusing on improving the system coverage, energy efficiency, or throughput, the works in \cite{Lyu2016}-\cite{Li2019} addressed UDUA problems for moving UAV-BSs.  Considering stable UAV-BSs, a centralized UAV-BS placement algorithm was proposed in \cite{Alzenad2018} to increase the number of covered ground users. Greedy method is a common selection while solving the user association problem, Hammouti et. al employed this as a benchmark in the UDUA solution \cite{El2021greedy}. The authors in \cite{Li2022iot} proposed a Dinkelbach based joint UDUA approach to maximize the energy efficiency. In order to reduce the total throughput of multiple UAV-BSs offloading mobile traffic from the terrestrial BSs, Zhang et al. \cite{Zhang2018} employed a Gaussian mixture model to predict the future traffic distribution in a considered area and then presented a gradient descent based UAV-BS deployment algorithm. In \cite{Mozaffari2016} and \cite{Mozaffari2017}, Mozaffari et al. proposed two iterative optimization based UDUA algorithms to minimize the downlink transmission power and uplink transmission power, respectively. Nevertheless, specific UDUA problems change temporally and spatially.
These model-driven UDUA approaches rely on iterative algorithms for the series of UDUA problems and introduce relatively long on-line computation time while lacking the computation power.

In the second category, machine learning technology has been incorporated into UDUA to cut down the problem's computational complexity. To improve the UAV-BSs' transmission energy efficiency, Liu et al. \cite{Liu2018} proposed a deep reinforcement learning based UAV deployment method where after being dispatched to the area of interest, the UAV-BSs gradually adjusted their positions according to their current statuses and channel conditions until they found their optimal positions. With the objective of maximizing the transmission rate of ground users or maximizing the system energy efficiency, two reinforcement learning based approaches were proposed in \cite{Mondal2021iot} and \cite{Abeywickrama2020}, respectively. The works in \cite{Liu2018}-\cite{Abeywickrama2020} allowed the UAV-BSs to explore and determine their proper positions after being dispatched, but signalling overhead between the UAV-BSs and the central controller was neglected. Moreover, the deep neural networks were trained in particular scenarios, which reflected that the mechanism needed to be re-trained for every new UDUA problem, so the time-efficiency was also degraded.

Our mechanism differs from the existing approaches in that 1) on-line time-efficiency is focused, a proper solution can be rapidly generated with the help of experiences accumulated from previous well-solved problems, 2)  it can be friendly transferred to new problems or scenarios, no extra training consumption is required. Thus, the system performance will be guaranteed and the on-demand response time can be much reduced.

\begin{figure}
    \centering
    \includegraphics[scale=0.9]{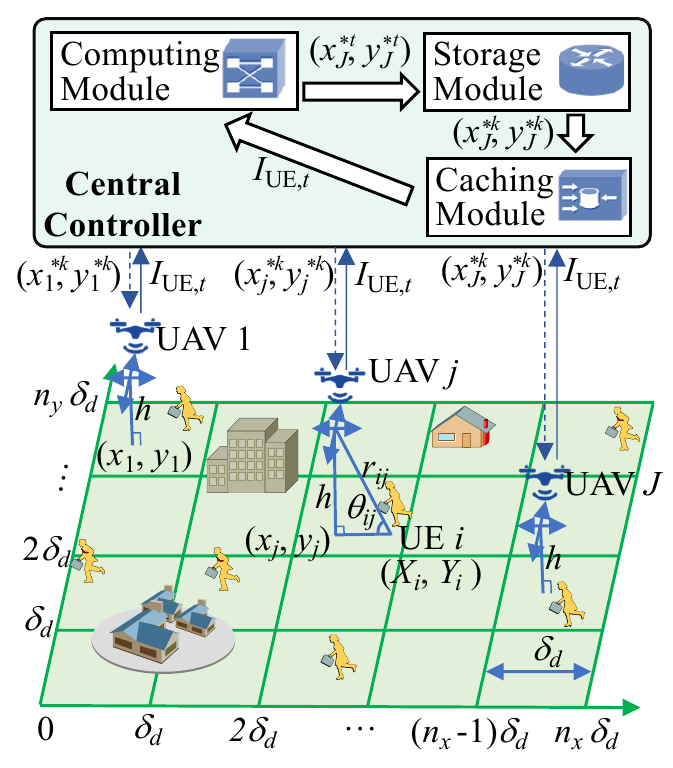}
    \caption{The system model.}
    \label{fig:system_model}
\end{figure}


\section{System Model}

This paper considers a UAV-aided communication system where multiple UAV-BSs are transported by vehicles to a certain region and requires to download data to ground user devices.
The system model is illustrated in Fig. \ref{fig:system_model}. We consider a classic and typical low-attitude UAV radio access network (RAN) scenario where $J$ UAV-BSs serve all the ground users located in a certain region $R$, and these UAV-BSs are controlled by a central controller, which is equipped with computing, caching, and storage modules. Since the downlink traffic is much higher than the uplink one in the usual multimedia communications \cite{Zhang2019}, this work only focuses on the downlink transmission.

Region $R$ is further divided into $n_y \times n_x$ grids with the same size of $\delta_d \times \delta_d$. We assume that $\delta_d$ is small enough so that different ground users in the same grid have the same channel condition with an arbitrary UAV-BS flying in the air \cite{luo2019two}. We also assume that the UAV RAN works in time intervals and the central controller has a global information.
At the beginning of every time interval, the central controller will first collect knowledge about ground user distribution in the $n_y \times n_x$ grids and then calculate the optimal UDUA strategy in a centralized way. Furthermore, we consider a quasi-static environment where the ground user distribution is assumed to be fixed during an arbitrary time interval.

In our model, each UAV-BS possesses $\Phi$ orthogonal frequency division multiple access (OFDMA) sub-channels, each of which has a fixed bandwidth of $B$. During a certain time interval, a UAV-BS can construct a downlink transmission connection with transmission power $p_t$ for one ground user with every sub-channel, and a ground user can be served by at most one UAV-BS. Taking advantage of proper spectrum management \cite{lopez2009ofdma}, we assume that the inter-UAV interference is well controlled and thus can be neglected. The influence of interference will be 
investigated in future work. We also assume that ground UEs in the region $R$ have the same external-interference condition with a constant noise power of $\sigma_n^2$ for analytical tractability. Being dispatched, the $J$ UAV-BSs will hover in fixed positions with the flight altitude of $h$.

We use sets $I _{\mathrm{UE},t}=\{\mathrm{UE}_1,\mathrm{UE}_2,...,\mathrm{UE}_I\}$ and $J _{\mathrm{UAV}}=\{\mathrm{UAV}_1,\mathrm{UAV}_2,...,\mathrm{UAV}_J\}$ to represent the set of ground users in region $R$ at time interval $t$, and the set of UAV-BSs, respectively. To guarantee that all the UEs can be served, we assume $I \leq J   \Phi$. $X_i$ and $Y_i$ are denoted as the ordinal numbers of $\mathrm{UE}_i$'s ($\mathrm{UE}_i \in I _{\mathrm{UE},t}$) position grid in latitude direction and longitude direction, respectively. Taking into account the fact that UAV-BSs are generally utilized in scenarios like IoT data transferring and crowd serving, this paper guarantees the basic quality-of-service (QoS) for ground users with the minimum data rate requirement of $C$. For each UAV-BS $\mathrm{UAV}_j \in J _{\mathrm{UAV}}$, two variables $x_j$ and $y_j$ are used to denote the grid location of its ground projection, and the Boolean variable $\delta_{ij}$ is used to denote its association relationship with ground user $i$ ($\delta_{ij}=1$ if ground user $i$ is served by UAV-BS $j$, $\delta_{ij}=0$ otherwise). This paper assumes that all the UAV-BSs only hover over region $R$ ($0 \le x_j \le n_x$ and $0 \le y_j \le n_y$ for $\forall j$).

According to \cite{al2014optimal}, the transmission channel between $\mathrm{UE}_i$ with position ($X_i$,$Y_i$) and UAV-BS $j$ will either have line-of-sight (LoS) propagation path or not when $x_j$ and $y_j$ are given depending on whether there are obstacles. Following \cite{al2014optimal}, the probability of UAV-BS $j$ having LoS propagation path with $\mathrm{UE}_i$ for certain $x_j$ and $y_j$ is calculated as:
\begin{equation}
    P^{\mathrm{LoS}}_{ij}= \frac{1}{1+a\;\mathrm{exp}(-b(\frac{180}{\pi}\mathrm{arcsin}(h/r_{ij}(x_j,y_j))-a))},
    \label{eq:probability_of_LoS}
\end{equation}
\noindent where $r_{ij}(x_j,y_j)$ is the 3-dimensional distance between $\mathrm{UE}_i$ and UAV-BS $j$, and $a$ and $b$ are constant parameters determined by the transmission environment.

The channel power gain between $\mathrm{UE}_i$ and UAV-BS $j$, ${g}_{ij}(x_j,y_j)$, is then calculated as:
\begin{equation}
  g_{ij}(x_j,y_j) =
    \begin{cases}
      \left(\frac{4\pi f}{c}\right)^{-2}\cdot r^{-\gamma}_{ij}(x_j,y_j)\cdot10^{-0.1\mu^{\mathrm{LoS}}}, \text{if LoS}\\
      \left(\frac{4\pi f}{c}\right)^{-2}\cdot r^{-\gamma}_{ij}(x_j,y_j)\cdot10^{-0.1\mu^{\mathrm{NLoS}}}, \text{others,}
    \end{cases}    
    \label{eq:power_gain_watt}
\end{equation}
where $f$, with the unit of Hz, is the frequency of the carrier signal, $c$ is the speed of light, $\gamma$ represents the large-scale pathloss exponent, $\mu^{\mathrm{LoS}}$ and $\mu^{\mathrm{NLoS}}$ are constants representing the excessive loss for the transmission channel with LoS propagation path or without LoS propagation path, respectively. As the sub-channels used by UAV-BSs have a relatively narrow bandwidth and are adjacent in the frequency domain, this work approximately assumes that $f$ is a constant for all the sub-channels.

According to Shannon's theorem, the data rate (in bits per second) of $\mathrm{UE}_i$ is given by:
\begin{equation}
    {C_{i}} = \sum\limits_{{\mathrm{UAV}_j }\in {J_{\mathrm{UAV}}}}
    {\delta _{ij}}\cdot B \cdot {\log _2} \left(1 + \frac {  {p_\mathrm{T} }{g_{ij}}({x_j},{y_j})} {{{\sigma_n ^2}}}\right),
   \label{eq:2-throughput}
\end{equation}

\noindent
where ${p_\mathrm{T}}{{g_{ij}}({x_j},{y_j})}$ is the received transmission power level at $\mathrm{UE}_i$.
From (3), we can clearly find that the achievable data rate of all the ground users depends on not only the locations of the UAV-BSs but also the association relationship between the UAV-BSs and the ground users.  

In this work, we propose to jointly optimize variables $x_j$, $y_j$, and $\delta_{ij}$ ($\mathrm{UE}_i \in I _{\mathrm{UE},t}$, $\mathrm{UAV}_j \in J _{\mathrm{UAV}}$), with the objective of maximizing the system's downlink sum throughput considering the basic QoS requirement of each ground user. Mathematically, the optimization problem can be formulated as follows:

(\textbf{P1:})
\begin{equation}
\argmax_{x_j, y_j, \delta_{ij}} \;\{\underset{i}{\Sigma}C_{i} \}
\label{eq:problem}
\end{equation}
\begin{equation}
    s.t\hspace{0.4cm}C1: \delta_{ij}=\{0,1\}, \forall i,j,
    \label{eq:connection01}
\end{equation}
\begin{equation}
\hspace{0.1cm}C2: \underset{j}{\Sigma}\delta_{ij}=1, \; \forall i,
    \label{eq:ue_unique_connection}
\end{equation}
\begin{equation}
  \hspace{0.2cm}  C3: \underset{i}{\Sigma} \;\delta_{ij}\leq \Phi, \; \forall j,
    \label{eq:uav_capacity}
\end{equation}
\begin{equation}
\hspace{-0.2cm}C4: C_{i} \geq C, \forall i,
\label{eq:data_rate_constraint}
\end{equation}
\begin{equation}
\begin{array}{c}
  \hspace{0.1cm} C5: 0\leq x_j \leq n_x, \\
  \hspace{0.8cm} 0\leq y_j \leq n_y, \\
 \hspace{1cm}  x_j\in \mathbb{Z}^+,y_j\in \mathbb{Z}^+.
\end{array}
\label{eq:uav_location_constraints}
\end{equation}

The problem (\ref{eq:problem}) is a classic joint optimization problem for maximizing the downlink sum throughput of the considered system. Even though this INLP problem can be solved, improving the time efficiency to meet the time-sensitive UAV services is still challenging. Constraint C1 (\ref{eq:connection01}) shows that $\delta_{ij}$ is a binary to control the set-up of connections. Constraint C2 (\ref{eq:ue_unique_connection}) ensures that any ground user $i$ is allowed to connect to only one UAV-BS at a time. 
Constraint C3 (\ref{eq:uav_capacity}) shows that the number of ground users served by a UAV-BS should be limited by the number of sub-channels it has.
Constraint C4 (\ref{eq:data_rate_constraint}) requires the basic QoS requirement of each ground user should be achieved. Finally, constraint C5 (\ref{eq:uav_location_constraints}) limits the hovering range of the UAV-BSs. A time-efficient solution for this problem will be typical while dealing with other related problems.

\vspace{-0.2cm}

\section{Proposed UDUA Mechanism}
From (\ref{eq:problem}), we can see that the user association can be performed when the UAV-BSs' locations are determined. In this section, we decouple the original optimization problem into the user association sub-problem and the UAV-BS deployment sub-problem. We also propose algorithms to solve these two sub-problems, respectively.


\vspace{-0.0cm}

\subsection{Decoupling P1}

By dividing variables $x_j$, $y_j$, and $\delta_{ij}$ ($\mathrm{UE}_i \in I _{\mathrm{UE},t}$, $\mathrm{UAV}_j \in J _{\mathrm{UAV}}$) into two groups, the original optimization problem of P1 can be decoupled into two sub-problems. One is the user association sub-problem which acquires the optimal matching strategy between the UAV-BSs and the ground users for given UAV-BS positions. The other is the UAV-BS deployment sub-problem, which tries to find the best position combination of the $J$ UAV-BSs making the first sub-problem's solution maximal among all the possible position combinations.

When positions of UAV-BSs are fixed ($x_j=\bar{x}_{j}, y_j=\bar{y}_{j}, \; \forall \mathrm{UAV}_j \in J _{\mathrm{UAV}}$), variables $\delta_{ij}$ ($\mathrm{UE}_i \in I _{\mathrm{UE},t}$, $\mathrm{UAV}_j \in J _{\mathrm{UAV}}$) will determine how the ground users are associated to the $J$ UAV-BSs. The user association sub-problem can be formulated as follows:

(\textbf{P1-1:})\begin{equation}
\argmax_{\delta_{ij} }\;\{\underset{i}{\Sigma} \bar{C}_{i}\}
\label{eq:problem1_1}
\end{equation}
\begin{equation}
s.t \; \; C1-C4,
\end{equation}

\noindent where constraints C1-C4 are defined in (\ref{eq:connection01})-(\ref{eq:data_rate_constraint}), and $\bar{C}_i$ for $\mathrm{UE}_i \in I_{\mathrm{UE},t}$ is calculated as:
\begin{equation}
    {\bar{C}_{i}} = \sum\limits_{{\mathrm{UAV}_j }\in {J_{\mathrm{UAV}}}}
    {\delta _{ij}}\cdot B \cdot {\log _2} \left(1 + \frac {  {p_\mathrm{T} }{g_{ij}}({\bar{x}_j},{\bar{y}_j})} {{{\sigma_n ^2}}}\right)
   \label{eq:2-avgthroughput}
\end{equation}

For given ground user set $I_{\mathrm{UE},t}$ and position combination of the $J$ UAV-BSs, i.e., $(\bar{x}_{1},...,\bar{x}_{J})$ and $(\bar{y}_{1},...,\bar{y}_{J})$, we define the optimal value of \textbf{P1-1} in (\ref{eq:problem1_1}) as $f_{I_{\mathrm{UE},t}}(\bar{x}_{1},...,\bar{x}_{J},\bar{y}_{1},...,\bar{y}_{J})$. Obviously, $f_{I_{\mathrm{UE},t}}({x}_{1},...,{x}_{J},{y}_{1},...,{y}_{J})$ can be seen as a function about the variables $x_j$ and $y_j$ ($\mathrm{UAV}_j \in J _{\mathrm{UAV}}$). Thus, the UAV-BS deployment sub-problem is formulated as:

(\textbf{P1-2:})
\begin{equation}
\argmax_{x_{j},y_j}\; f_{I_{\mathrm{UE},t}}({x}_{1},...,{x}_{J},{y}_{1},...,{y}_{J})
\label{eq:problem1-2}
\end{equation}
\begin{equation}
s.t \; \; C5,
\end{equation}
\noindent where constraint C5 is defined by (\ref{eq:uav_location_constraints}).

\vspace{-0.0cm}

\subsection{Solution for the User Association Sub-problem}

When position of UAV-BS $j$ is given as ($\bar{x}_{j}$, $\bar{y}_{j}$), the channel pathloss between UAV-BS $j$ and ground user $i$ has the certain value of ${g_{ij}}(\bar{x}_{j}, \bar{y}_{j})$ according to (\ref{eq:power_gain_watt}). If user $i$ is matched with UAV-BS $j$, we use $C_{ij}(\bar{x}_{j}, \bar{y}_{j})$ to present the achievable transmission data rate as follows:
\begin{equation}
    {C_{ij}(\bar{x}_{j}, \bar{y}_{j})} =  B \cdot {\log _2} \left(1 + \frac {  {p_\mathrm{T} }{g_{ij}}({\bar{x}_j},{\bar{y}_j})} {{{\sigma_n ^2}}}\right)
   \label{eq:2-singthroughput}
\end{equation}

Obviously, if $C_{ij}(\bar{x}_{j}, \bar{y}_{j}) <C$, user $i$ can not be associated to UAV-BS $j$ due to the minimum data rate constraint (\ref{eq:data_rate_constraint}). Otherwise, UAV-BS $j$ can serve user $i$.

\begin{figure}
    \centering
    \includegraphics[scale=0.45]{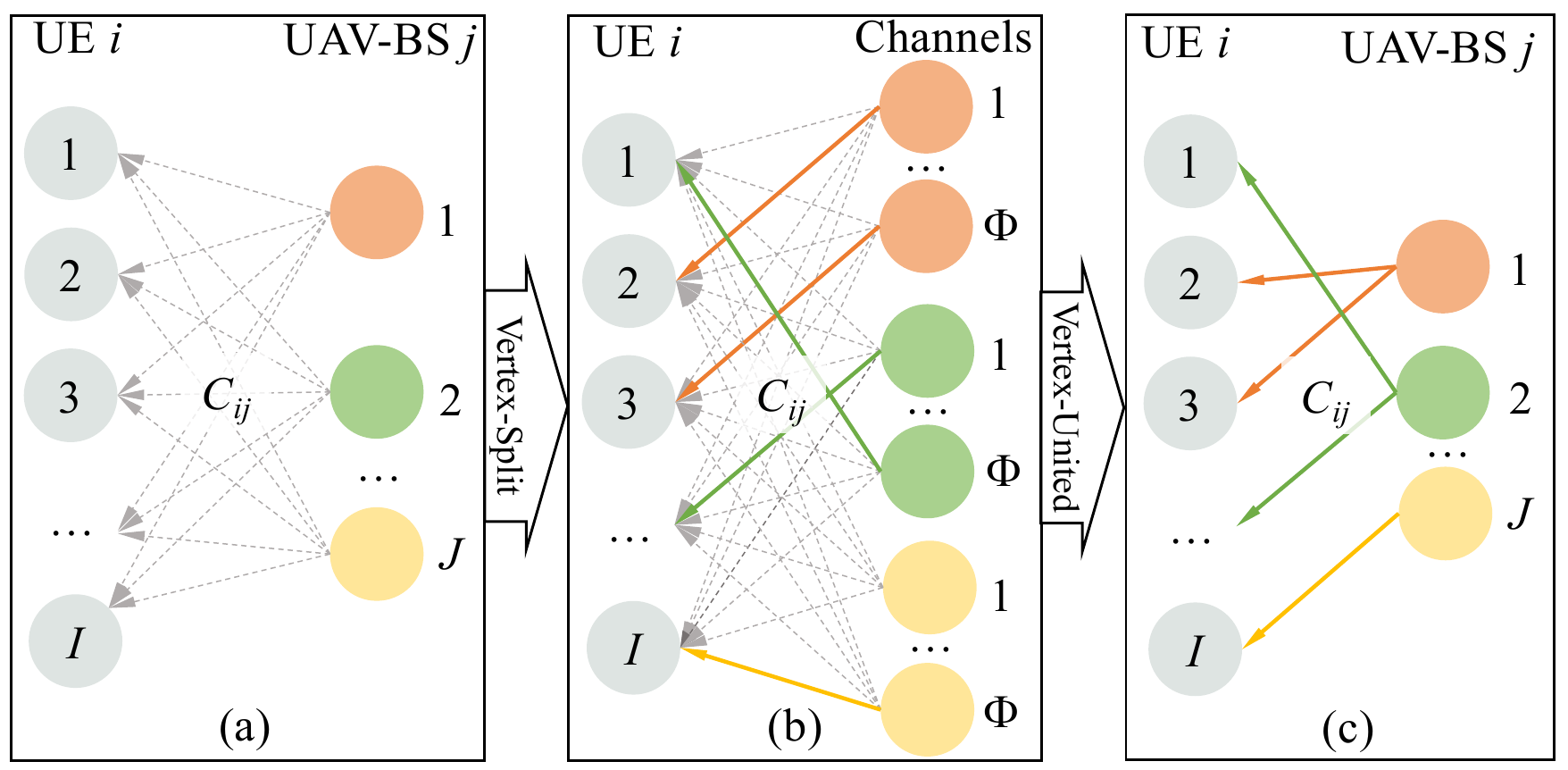}
    \caption{Node-split KM algorithm to allocate UEs to UAV-BSs with the capacity threshold.}
    \label{fig:nodesplit_KM}
\end{figure}

We represent the $J$ UAV-BSs and the $I$ ground users as two groups of vertexes shown in Fig. \ref{fig:nodesplit_KM}(a). For the vertex related to user $i$ and the vertex related to UAV-BS $j$, they will have a link with weight $C_{ij}(\bar{x}_{j}, \bar{y}_{j})$ as long as $C_{ij}(\bar{x}_{j}, \bar{y}_{j}) \ge C$, and can not connect to each other once $C_{ij}(\bar{x}_{j}, \bar{y}_{j}) < C$. Then, the user association sub-problem P1-1 is equivalent to a coloring problem for a bipartite graph, where the objective is to maximize the sum weight of the colored links and the following principles should be satisfied:

1) The link between ground user $i$ and UAV-BS $j$ is colored when and only when user $i$ is served by UAV-BS $j$ ($\delta_{ij}=1$);

2) In accordance with C2 (\ref{eq:ue_unique_connection}) that a ground user must be served by one UAV-BS in set $J$, the vertex related to any user will have and only have one colored link to the vertexes related to the UAV-BSs;

3) In accordance with C3
(\ref{eq:uav_capacity}) that a UAV-BS will at most serve $\Phi$ ground users due to its limited OFDMA sub-channels, not more than $\Phi$ colored links can be connected to the vertex related to any UAV-BS in set $J$.

We split every vertex related to a UAV-BS in Fig. \ref{fig:nodesplit_KM}(a) into $\Phi$ vertexes as shown in Fig. \ref{fig:nodesplit_KM}(b). The links between each of the $\Phi$ vertexes related to UAV-BS $j$ and the ground user vertexes in Fig. \ref{fig:nodesplit_KM}(b) have the same weight values as those between the original UAV-BS vertex and the ground user vertexes. Formally, if ground user $i$ can not be associated to UAV-BS $j$ in Fig. \ref{fig:nodesplit_KM}(a), we set links between the user vertex and the split UAV-BS vertexes in Fig. \ref{fig:nodesplit_KM}(b) with a constant negative weight, $-W$, whose absolute value is much larger than $C$ ($W>>C$). Thus, the coloring problem in Fig. \ref{fig:nodesplit_KM}(a) can further be transformed into the coloring problem in Fig. \ref{fig:nodesplit_KM}(b) with the same objective of maximizing the sum weight of the colored links. Different from Fig. \ref{fig:nodesplit_KM}(a), each split UAV-BS vertex in Fig. \ref{fig:nodesplit_KM}(b) can have one colored link to the user vertexes at most. The coloring problem in Fig. \ref{fig:nodesplit_KM}(b) is a typical maximum-weight one-to-one matching problem of a bipartite graph, which can be solved efficiently by the existing Kuhn-Munkres algorithm [\ref{fig:nodesplit_KM}]. We should note that the equivalent user association sub-problem will have no feasible solution when there is at least one negative-weight link in Fig. \ref{fig:nodesplit_KM}(b) being colored by the Kuhn-Munkres algorithm.

Finally, as illustrated in Fig. \ref{fig:nodesplit_KM}(c), all the ground users possessing a colored link to the split UAV-BS vertexes related to UAV-BS $j$ will be associated to this UAV-BS. The optimal value of P1-1, $f_{I_{\mathrm{UE},t}}(\bar{x}_{1},...,\bar{x}_{J},\bar{y}_{1},...,\bar{y}_{J})$, can also be obtained by adding the weights of colored links ($C_{ij}(\bar{x}_{j}, \bar{y}_{j})$) together if it has feasible solutions. Also, when P1-1 does not have feasible solutions for a certain UAV-BS deployment strategy, we formally record $f_{I_{\mathrm{UE},t}}({x}_{1},...,{x}_{J},{y}_{1},...,{y}_{J})$ as $- I \times W$.

\vspace{-0.0cm}

\subsection{Solution for the UAV-BS Deployment Sub-problem}

Based on the solution of P1-1 for any given UAV-BS deployment strategy, we can use the exhaustive searching approach to test all the possible location combinations of the considered UAV-BSs and choose the best one that achieves the maximum $f_{I_{\mathrm{UE},t}}({x}_{1},...,{x}_{J},{y}_{1},...,{y}_{J})$ value. Nevertheless, this exhaustive searching approach is not proper for on-line UDUA problems since the searching space augments exponentially as the UAV-BS number gets large. For $n_y \times n_x$ grids and $J$ UAV-BSs considered, there are $(n_y \times n_x)^J$ possible UAV-BS deployment strategies in summary.

To reduce the computation complexity, this paper solves the UAV-BS deployment sub-problem by imitating the way of thinking used by humans. Inspired by the phenomenon that people tend to handle a new problem utilizing the experiences and knowledge from previously solved ones, we analyze whether the optimal UAV-BS deployment strategies of given ground user distributions can help to provide a proper UAV-BS deployment strategy for any newly considered ground user distribution. 

\textbf{Lemma 1: }
We use $I _{\mathrm{UE},1}$ to represent an arbitrary set of ground users, and use $({x}_{1},...,{x}_{J},{y}_{1},...,{y}_{J})$ to represent a certain deployment strategy of the $J$ UAV-BSs. For any ground user ${u_a} \notin {I _{\mathrm{UE},1}}$, we use $I _{\mathrm{UE},2}$ to represent ${I _{\mathrm{UE},1}} \cup \left\{ {{u_a}} \right\}$. If the UAV-BS deployment strategy $({x}_{1},...,{x}_{J},{y}_{1},...,{y}_{J})$ makes the user association sub-problems related to both $I _{\mathrm{UE},1}$ and $I _{\mathrm{UE},2}$ have feasible solutions, then for an arbitrary feasible user association strategy of $I _{\mathrm{UE},1}$, we can connect ${u_a}$ to a proper UAV-BS with available sub-channels by adjusting the connecting statuses of up to $J-1$ ground users in $I _{\mathrm{UE},1}$.

\textit{Proof:}
See Appendix A.

We can further prove \textbf{Lemma 2}.

\textbf{Lemma 2: }
For a given set of ground users, $I _{\mathrm{UE},1}$, we use $({x}^{*1}_{1},...,{x}^{*1}_{J},{y}^{*1}_{1},...,{y}^{*1}_{J})$ to represent the optimal UAV-BS deployment strategy related to $I _{\mathrm{UE},1}$. Then for an arbitrary set of ground users, $I _{\mathrm{UE},2}$, where $m$ new ground users are added to $I _{\mathrm{UE},1}$, if $({x}^{*1}_{1},...,{x}^{*1}_{J},{y}^{*1}_{1},...,{y}^{*1}_{J})$ makes the user association sub-problem of $I _{\mathrm{UE},2}$ have feasible solutions, we obtain the following inequality:
\begin{equation}
\begin{array}{l}
{f_{{I _{\mathrm{UE},2}}}}(x_1^{*1},\;...,\;x_J^{*1},\;y_1^{*1},\;...,\;y_J^{*1}) \ge \\
{f_{{I _{\mathrm{UE},1}}}}(x_1^{*1},\;...,\;x_J^{*1},\;y_1^{*1},\;...,\;y_J^{*1}) + \\
m  [{\epsilon _{\mathrm{min} }} - (J - 1)  ({\epsilon _{\mathrm{max} }} - {\epsilon _{\mathrm{min} }})],
\end{array}
\label{eq:lemma2-21}
\end{equation}

\noindent
where ${f_{{I _{\mathrm{UE},1}}}}(x_1^{*1},\;...,\;x_J^{*1},\;y_1^{*1},\;...,\;y_J^{*1})$ and ${f_{{I _{\mathrm{UE},2}}}}(x_1^{*1},\;...,\;x_J^{*1},\;y_1^{*1},\;...,\;y_J^{*1})$ are the optimal values of the user association sub-problems related to $I _{\mathrm{UE},1}$ and $I _{\mathrm{UE},2}$, respectively, when the UAV-BS deployment strategy is $({x}^{*1}_{1},...,{x}^{*1}_{J},{y}^{*1}_{1},...,{y}^{*1}_{J})$, ${\epsilon _{\mathrm{max} }}$ is the maximum data rate that can be achieved by a UAV-BS to serve a ground user in the considered system, and ${\epsilon _{\mathrm{min} }}$ is the minimum data rate required by a UE.

\textit{Proof:}
See Appendix B. 

With \textbf{Lemma 1} and \textbf{Lemma 2}, \textbf{Proposition 1} can be proved.

\textbf{Proposition 1:}
We use $({x}^{*1}_{1},...,{x}^{*1}_{J},{y}^{*1}_{1},...,{y}^{*1}_{J})$ and $({x}^{*2}_{1},...,{x}^{*2}_{J},{y}^{*2}_{1},...,{y}^{*2}_{J})$ to represent the optimal UAV-BS deployment strategies for two given sets of ground users, $I _{\mathrm{UE},1}$ and $I _{\mathrm{UE},2}$, respectively. If we can get $I _{\mathrm{UE},2}$ by adding $m$ ground users into or removing $m$ ground users off $I _{\mathrm{UE},1}$, and $({x}^{*1}_{1},...,{x}^{*1}_{J},{y}^{*1}_{1},...,{y}^{*1}_{J})$ and $({x}^{*2}_{1},...,{x}^{*2}_{J},{y}^{*2}_{1},...,{y}^{*2}_{J})$ both make the user association sub-problems related to $I _{\mathrm{UE},1}$ or $I _{\mathrm{UE},2}$ have feasible solutions, then we can arrive at the following inequality:
\begin{equation}
\begin{array}{l}
{f_{{I _{\mathrm{UE},2}}}}(x_1^{*1},\;...,\;x_J^{*1},\;y_1^{*1},\;...,\;y_J^{*1}) \ge \\
{f_{{I _{\mathrm{UE},2}}}}(x_1^{*2},\;...,\;x_J^{*2},\;y_1^{*2},\;...,\;y_J^{*2}) -
m  J   ({\epsilon _{\mathrm{max}}} - {\epsilon _{\mathrm{min}}}).
\end{array}
\label{eq:proposition1-22}
\end{equation}

\textit{Proof:}
See Appendix C.

From \textbf{Proposition 1}, we can conclude that, under certain conditions, adopting the optimal UAV-BS deployment strategy of a previous ground user set for a new ground user set will introduce limited downlink sum throughput reduction compared with this new set's own optimal UAV-BS deployment strategy, if the new user set is achieved by adding some ground users into or removing some ground users off the previous set. Also, the upper bound of this reduced downlink sum throughput for the new ground user set is linearly correlated to the user number difference between the two ground user sets.

For a given ground user set, $I _{\mathrm{UE},1}$, when there are ground users moving inside the considered region $R$, we can prove the following \textbf{Lemma 3}.

\textbf{Lemma 3: }
For an given set of ground users, $I _{\mathrm{UE},1}$, we use $({x}^{*1}_{1},...,{x}^{*1}_{J},{y}^{*1}_{1},...,{y}^{*1}_{J})$ to represent the optimal UAV-BS deployment strategy related to $I _{\mathrm{UE},1}$. If $({x}^{*1}_{1},...,{x}^{*1}_{J},{y}^{*1}_{1},...,{y}^{*1}_{J})$ makes the ground user set $I _{\mathrm{UE},2}$, where $n$ ground users in $I _{\mathrm{UE},1}$ change their position grids, have feasible solutions for the corresponding user association problem, we will get the following relationship:
\begin{equation}
\begin{array}{*{20}{l}}
{{f_{{I _{\mathrm{UE},2}}}}(x_1^{*1},\;...,\;x_J^{*1},\;y_1^{*1},\;...,\;y_J^{*1}) \ge }\\
{{f_{{I _{\mathrm{UE},1}}}}(x_1^{*1},\;...,\;x_J^{*1},\;y_1^{*1},\;...,\;y_J^{*1}) - }
{n  J  ({\epsilon_{\mathrm{max}}} - {\epsilon_{\mathrm{min}}})}.
\end{array}
\label{eq:lemma3-23}
\end{equation}

\textit{Proof:}
See Appendix D. 

Based on \textbf{Lemma 3}, we can also prove \textbf{Proposition 2}.

\textbf{Proposition 2:}
We use $({x}^{*1}_{1},...,{x}^{*1}_{J},{y}^{*1}_{1},...,{y}^{*1}_{J})$ and $({x}^{*2}_{1},...,{x}^{*2}_{J},{y}^{*2}_{1},...,{y}^{*2}_{J})$ to represent the optimal UAV-BS deployment strategies for two given sets of ground users, $I _{\mathrm{UE},1}$ and $I _{\mathrm{UE},2}$, respectively. $I _{\mathrm{UE},2}$ is acquired by changing the position grids of $n$ ground users in $I _{\mathrm{UE},1}$. If $({x}^{*1}_{1},...,{x}^{*1}_{J},{y}^{*1}_{1},...,{y}^{*1}_{J})$ and $({x}^{*2}_{1},...,{x}^{*2}_{J},{y}^{*2}_{1},...,{y}^{*2}_{J})$ both make the user association sub-problems related to $I _{\mathrm{UE},1}$ or $I _{\mathrm{UE},2}$ have feasible solutions, then we can arrive at the following inequality:
\begin{equation}
\begin{array}{*{20}{l}}
{{f_{{I _{\mathrm{UE},2}}}}(x_1^{*1},\;...,\;x_J^{*1},\;y_1^{*1},\;...,\;y_J^{*1}) \ge }\\
{{f_{{I _{\mathrm{UE},2}}}}(x_1^{*2},\;...,\;x_J^{*2},\;y_1^{*2},\;...,\;y_J^{*2}) - }
{2  n  J  ({\epsilon _{\mathrm{max}}} - {\epsilon _{\mathrm{min}}})}.
\end{array}
\label{eq:proposition2-24}
\end{equation}
\textit{Proof:}
See Appendix E.

Similar with \textbf{Proposition 1}, \textbf{Proposition 2} shows that under certain conditions, adopting the optimal UAV-BS deployment strategy of a previous ground user set for a new ground user set will introduce limited downlink sum throughput reduction compared with the new user set's own optimal UAV-BS deployment strategy, when the new user set can be achieved from the previous user set by moving some ground users inside the considered region $R$. Furthermore, the upper bound of this reduced downlink sum throughput for the new ground user set is proportional to the number of users moved.

\textbf{Proposition 1} and \textbf{Proposition 2} imply that even though the optimal UAV-BS deployment strategy of a previous ground user set, $I _{\mathrm{UE},1}$, isn't the best UAV-BS deployment strategy of a new user set, $I _{\mathrm{UE},2}$, adopting this UAV-BS deployment strategy for $I _{\mathrm{UE},2}$ is likely to introduce limited downlink sum throughput reduction compared with $I _{\mathrm{UE},2}$'s actual optimal UAV-BS deployment strategy if these two ground user sets are similar ($m$ and $n$ in (\ref{eq:proposition1-22}) or (\ref{eq:proposition2-24}) are small). From (\ref{eq:proposition1-22}) and (\ref{eq:proposition2-24}), we also see that each ground user moved inside the considered region seems to have a double effect on the upper bound of this reduction than a user moved in or out.

Based on \textbf{Proposition 1} and \textbf{Proposition 2}, we define the difference degree between two ground user sets and propose a KNN \cite{cover1967nearest} based algorithm to solve the UAV-BS deployment sub-problem. We use an ${n_y} \times {n_x}$ matrix $D_t$ to represent the user distribution of a certain ground user set, $I _{\mathrm{UE},t}$. Each element $D_t(k_y,k_x)$ is an integer which records the number of ground users in $I _{\mathrm{UE},t}$ located in grid $(k_y,k_x),{k_y} = 1,...,{n_y},{k_x} = 1,...,{n_x}$. For two ground user sets $I _{\mathrm{UE},1}$ and $I _{\mathrm{UE},2}$, we define their difference degree as follows.

\begin{figure}
    \centering
    \includegraphics[scale=0.8]{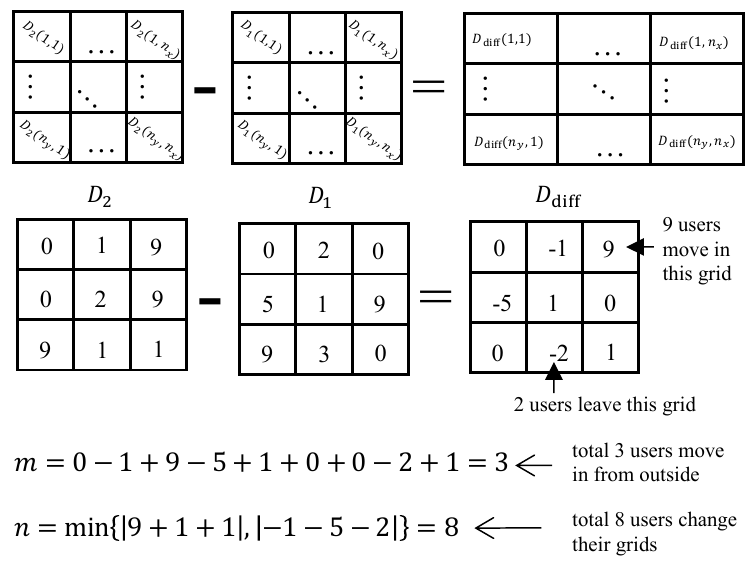}
    \caption{Illustration of calculating the difference matrix $D_{\mathrm{diff}}$ and the parameters of difference degree, $m$ and $n$.}
    \label{fig:m_n_matrix}
\end{figure}

\textbf{Definition:} For two arbitrary ground user sets $I _{\mathrm{UE},1}$ and $I _{\mathrm{UE},2}$, we obtain their difference matrix, $D_{\mathrm{diff}}$, by operating the matrix subtraction between user distribution matrices related to the two user sets, $D_1$ and $D_2$, as shown in Fig. \ref{fig:m_n_matrix}. The difference degree between $I _{\mathrm{UE},1}$ and $I _{\mathrm{UE},2}$ is defined as:
\begin{equation}
\Gamma_{\mathrm{diff}}(I _{\mathrm{UE},1},I _{\mathrm{UE},2}) = m+ 2  n,
\label{eq:difference_degree}
\end{equation}

\noindent
where $m$ and $n$ denote, compared to $I _{\mathrm{UE},1}$, the number of ground users in $I _{\mathrm{UE},2}$ moved in or moved out of the considered region and the number of ground users in $I _{\mathrm{UE},2}$ moved inside the considered region, respectively. $m$ and $n$ can be calculated through $D_{\mathrm{diff}}$:
\begin{equation}
m = \left|\sum\limits_{\substack{{k_y} = 1,...,{n_y},\\{k_x} = 1,...,{n_x}}} {{D_{\mathrm{diff}}}({k_y},{k_x})} \right|,
\label{eq:calculate_m}
\end{equation}
\begin{equation}
\begin{array}{l}
n = \mathrm{min} \Bigg\{ 
\sum\limits_{\substack{{k_y} = 1,...,{n_y},\\{k_x} = 1,...,{n_x},\\{D_{\mathrm{diff}}}({k_y},{k_x}) > 0}} {|{D_{\mathrm{diff}}}({k_y},{k_x})|} ,\\
\hspace{1.5cm}\sum\limits_{\substack{ {k_y} = 1,...,{n_y},\\{k_x} = 1,...,{n_x},\\{D_{\mathrm{diff}}}({k_y},{k_x}) < 0}} {|{D_{\mathrm{diff}}}({k_y},{k_x})} |  \Bigg\} .
\end{array}
\label{eq:calculate_n}
\end{equation}

\begin{figure}
    \centering
    \includegraphics[scale=0.78]{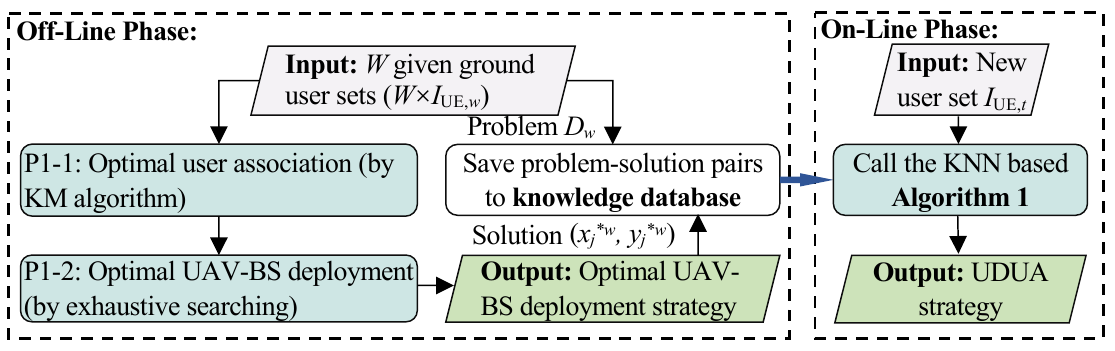}
    \caption{Process description of the proposed algorithm.}
    \label{fig:generalframework}
\end{figure}

The solution to the second sub-problem is demonstrated in Fig. \ref{fig:generalframework}. At the off-line phase, the proposed algorithm stores the optimal UAV-BS deployment strategies of $W$ given ground user sets to construct a knowledge database in advance. This knowledge database can be viewed as an analogy to a human's experience, which we use to handle the new problems. For each ground user set $I _{\mathrm{UE},w}$ in the knowledge database, we use matrix $D_w$ to record the user distribution and get its optimal UAV-BS deployment strategy $({x}^{*w}_{1},...,{x}^{*w}_{J},{y}^{*w}_{1},...,{y}^{*w}_{J})$ by exhaustively comparing all the $(n_y \times n_x)^J$ possible UAV-BS deployment strategies. Notably, although preparing the knowledge database is relatively computing-resource consuming, we can accomplish this task before the UAV RAN is set, and thus it will not influence the running time of each on-line UDUA problem. For each newly considered UDUA problem with ground user set $I _{\mathrm{UE},t}$, the proposed UAV-BS deployment algorithm will first calculate $I _{\mathrm{UE},t}$'s difference degree to each ground user set in the knowledge database. Then, the proposed algorithm will compare the optimal UAV-BS deployment strategies related to the $k$ ground user sets in the knowledge database, which have the smallest difference degrees with $I _{\mathrm{UE},t}$, and select the feasible one achieving the maximum downlink throughput for $I _{\mathrm{UE},t}$. The pseudo code of our UAV-BS deployment algorithm's on-line phase is given in \textbf{Algorithm 1}.

\begin{algorithm}[t]
\SetAlgoLined
    \SetKwData{Left}{left}\SetKwData{This}{this}\SetKwData{Up}{up}
    \SetKwFunction{Union}{Union}\SetKwFunction{FindCompress}{FindCompress}
    \SetKwInOut{Input}{input}\SetKwInOut{Output}{output}
    \Input{a new ground-user set $I _{\mathrm{UE},t}$, the knowledge database: given ground-user sets $W\times I _{\mathrm{UE},w}$ and associated UAV-BS deployment strategies $W\times({x}^{*w}_{1},...,{x}^{*w}_{J},{y}^{*w}_{1},...,{y}^{*w}_{J})$.}
    \Output{UAV-BS deployment $({x}^{*k}_{1},...,{x}^{*k}_{J},{y}^{*k}_{1},...,{y}^{*k}_{J})$ and user association strategy of $I _{\mathrm{UE},t}$.}
    generate the distribution matrix $D_t$ according to $I _{\mathrm{UE},t}$\;
    \For{\normalfont each user set matrix $D$ in $W$}{
        calculate $m$ and $n$ with $D_t$ and $D$ based on (\ref{eq:calculate_m}) and (\ref{eq:calculate_n})\;
        generate the difference degree $\Gamma_{\mathrm{diff}}$ (\ref{eq:difference_degree})\;
    }
    select the top $K$ ground user sets from the knowledge database processing the minimum $\Gamma_{\mathrm{diff}}$ with $D_t$\;
    
    \For{\normalfont each ground user set $D_k$ of the $K$ selected ones}{
        retrieve the optimal UAV-BS deployments strategy related to $D_k$, $({x}^{k}_{1},...,{x}^{k}_{J},{y}^{k}_{1},...,{y}^{k}_{J})$\;
        run the \textbf{Kuhn-Munkres} on $\{I _{\mathrm{UE},t},({x}^{k}_{1},...,{x}^{k}_{J},{y}^{k}_{1},...,{y}^{k}_{J})\}$\;
        \uIf {\normalfont Kuhn-Munkres has a feasible solution} {
            record the values of throughput $f_{D_t}({x}^{k}_{1},...,{x}^{k}_{J},{y}^{k}_{1},...,{y}^{k}_{J})$\;
        }
        \Else{
        \Output{no feasible solution}
        }
    }
    get the UAV-BS deployment solution $({x}^{*k}_{1},...,{x}^{*k}_{J},{y}^{*k}_{1},...,{y}^{*k}_{J})$ with the maximum throughput\;
    \Output{$({x}^{*k}_{1},...,{x}^{*k}_{J},{y}^{*k}_{1},...,{y}^{*k}_{J})$ and the related user association strategy}
 \caption{KNN based UDUA algorithms in on-line phase}
\end{algorithm}

\vspace{-0.0cm}
\subsection{Computational Complexity of An On-line UDUA Problem}
For an on-line UDUA problem with ground user set $I _{\mathrm{UE},t}$, constructing its user distribution matrix $D_t$ has the complexity of $O(I)$, where $I$ is the number of ground users; calculating the difference matrices and difference degrees between $I _{\mathrm{UE},t}$ and the $W$ given ground user sets both have the complexity of $O(W  {n_y}  {n_x})$, where ${n_y} \times {n_x}$ are the total grid number of the considered region; finding the $k$ ground user sets in the knowledge database possessing the smallest difference degrees with $I _{\mathrm{UE},t}$ has the complexity of $O(W)$. In line 9 of \textbf{Algorithm 1}, solving the user-association sub-problem for $I _{\mathrm{UE},t}$ with the UAV-BS deployment strategy related to each of the $k$ selected ground user sets using the Kuhn-Munkres algorithm has the complexity of $O(I^{4})$ \cite{kuhn1955hungarian}. Finally, choosing the feasible UAV-BS deployment strategy, which achieves the maximum downlink throughput for $I _{\mathrm{UE},t}$ among the $k$ candidate ones has the complexity of $O(k)$. Thus, the overall computational complexity of an on-line UDUA problem is bounded by $O(W  {n_y}  {n_x} + {I^4}  k)$.

Notably, for a candidate UAV-BS deployment strategy and a considered ground user set $I _{\mathrm{UE},t}$, the channel power gain between each UAV-BS and each ground user can be acquired directly by reading a table that provides all the possible channel power gain values between a UAV-BS and a ground user when they are located in the rasterised region $R$. As a result, we do not take the complexity of calculating these channel power gains into consideration in our complexity analysis.

\;

\vspace{-1cm}

\section{Experimental Results}


We evaluate the performance of our UDUA mechanism through extensive experiments. In this section, our experimental settings are first described. Then, we test how the two key hyper-parameters, i.e., the scale of the knowledge database, $W$, and the number of candidate UAV-BS deployment strategies, $k$, will influence the proposed mechanism's performance. We also compare our UDUA mechanism with some baseline UDUA approaches under various network scenarios. Finally, experimental results about storage resources needed as well as the off-line and on-line computational time of our mechanism with different hyper-parameter values will be provided.

\vspace{-0.0cm}
\subsection{Experimental Parameters}
In our experiments, we consider a $90\;\mathrm{m} \times 90\;\mathrm{m}$ region and evenly divide it into $9 \times 9$ grids (${n_y} = {n_x} = 9$). The users distributions are simulated according to the findings in \cite{lee2014spatial} by Lee et al. that UEs are distributed non-uniformly, tending to gather together in some hot-spots, and requiring more communications resources than other areas. Specifically, we follow the work in \cite{lee2014spatial} and use a log-normal distribution with parameters $\mu$ and $\sigma$ to fit the number of ground users in each grid in the region $R$. $\mu$ and $\sigma$ jointly determine the density of ground users in $R$, and $\sigma$ denotes how non-uniformly the ground users are distributed. It should be noted that $\mu$ and $\sigma$ do not determine locations of hotspots, so the user distribution can be very different even with the same mean and variance. We vary the value of $\mu$ in set $\{-1,-0.8,-0.6,-0.4,-0.2\}$ and vary the value of $\sigma$ in set $\{0.2,0.4,0.6,0.8,1\}$. The user amount differs from tens to hundreds. Though the 25 value combinations of $\mu$ and $\sigma$ can not depict all the possible ground user distributions in the real world, they comprise lots of general RAN scenarios where the density and the non-uniformity of ground users range widely.

For each of the 25 value combinations of $\mu$ and $\sigma$, we randomly generate $W/25$ ground user sets to construct the knowledge database and use the exhaustive searching approach to obtain their optimal UAV-BS deployment strategies, which are denoted as the theoretical optimal (TO). We also randomly generate $N_{\mathrm{Test}}$ testing ground user sets related to every value combination of $\mu$ and $\sigma$ to evaluate the proposed UDUA mechanism's performance. In order to demonstrate the efficiency of our UDUA mechanism, we compare it with four kinds of baseline algorithms. The first one is a combination of exhaustive UAV-BS deployment and Kuhn-Munkres based user association, which offers the TO theoretical optimal results.
The second one is simulated annealing based UAV-BS deployment with greed algorithm based user association (SAUD-GUA) as the benchmark in the literature \cite{El2021greedy}. In SAUD-GUA, simulated annealing is a heuristic approach which sacrifices limited performance for reducing the time complexity and the greed algorithm solves user association by connecting ground users owning the best channel conditions first. SAUD-GUA is a common mode in literature with acceptable computational complexity. We also combine simulated annealing based UAV-BS deployment with Kuhn-Munkres based user association (SAUD-KMUA) as the third baseline algorithm for the comparison. The final baseline algorithm (RUD-GUA) uses the random approach, which randomly generates locations of UAV-BSs, for the UAV-BS deployment, and associates the ground users to the UAV-BSs with the greed algorithm. It is no doubt that RUD-GUA has the lowest time-complexity among all the considered UDUA approaches.

If an approach does not find a feasible UDUA solution for a specific testing ground user set, we will record one failure to this approach. The failure rate of a UDUA approach is calculated by the following equation:
\begin{equation}
\mathrm{Failure}\;\mathrm{rate} = \frac{{{N_{\mathrm{Fail}}}}}{{{N_{\mathrm{Test,Sum}}}}},
\end{equation}

\noindent
where ${{N_{\mathrm{Fail}}}}$ is the failure number of a UDUA approach and ${{N_{\mathrm{Test,Sum}}}}$ is the number of testing ground user sets.

In our experiments, the UAV-BSs are working in the hovering model with a fixed height of 20 m. The minimum required data rate for each ground user is set as 300 bps for the basic audio and video services. Our experiments are executed on a desktop computer with a 3.6 GHz AMD Ryzen 7 3700X processor and a 16GB LPDDR4 RAM. The values of major experimental parameters are summarized in \textbf{Table 1} according to 3GPP-LTE based RAN systems \cite{3gpp.36.777}.

\vspace{-0.0cm}

\subsection{Influence of Key Hyper-parameters on the Proposed UDUA Mechanism}
As described in \textbf{Algorithm 1}, the proposed UDUA mechanism compares the optimal UAV-BS deployment strategies related to the $k$ most similar ground user sets in the knowledge database with size $W$. As a result, both $W$ and $k$ are two key hyper-parameters that will have an influence on the proposed mechanism's performance.

\begin{table}[]
    \centering
    \scalebox{0.7}{
    \begin{tabular}{l l l}
    \hline
        Parameters &Description& Values  \\
        \hline
        $B$ & Sub-channel bandwidth &0.1 MHz\\
        $C$&Data-rate requirement&300 kb/s\\
        $\gamma$&Path loss exponent&3\\
        $f$&Frequency& 2 GHz\\
        $n_y\times n_x$&Total grids&81 \\ 
        $\delta_d$&Unit side length&10 m\\
        $h$&Height of UAV-BSs&20 m \\
        $N_{\mathrm{Test}}$ &Size of test user sets& 20 \\
        $p _{\mathrm{T}}$ &Transmission power&20 dBm\\
        $W$&Size of database&500 \\ 
        $J$&Amount of UAV-BSs&2 \\ 
        $\mu$&Log-normal parameter&\{-1,\;-0.8,\;-0.6,\;-0.4,\;-0.2\}\\
        $\sigma$&Log-normal parameter&\{0.2,\;0.4,\;0.6,\;0.8,\;1\}\\
        $\sigma_n^2$&Noise power& -125 dBm\\
        (a,b)&Pathloss model parameters (urban)& (9.6117,\;0.2782) \cite{al2014optimal}\\
        ($\mu^{\mathrm{LoS}}$,$\mu^{\mathrm{NLoS}}$)&Mean of additive pathloss (urban) &(1,20) \cite{al2014optimal}\\
        \hline
    \end{tabular}}
    \caption{Parameter values in experiment.}
    \label{tab:parameter}
\end{table}

\begin{figure}
    \centering
    \includegraphics[scale=0.5]{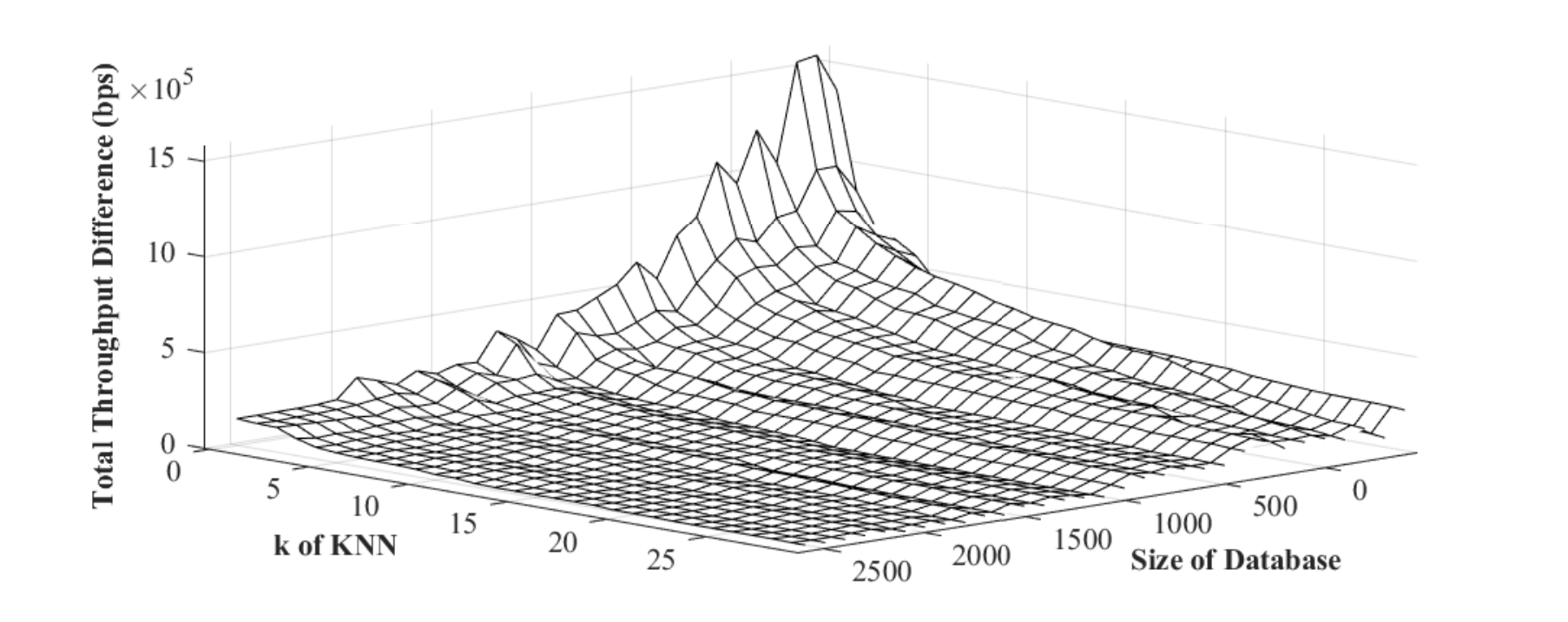}
    \caption{Estimated performance of the proposed UDUA with two key hyper-parameters $W$ and $k$.   }
    \label{fig:1_1_delta_p_k_Nued}
\end{figure}

Fig. \ref{fig:1_1_delta_p_k_Nued} presents the performance gap between the proposed UDUA mechanism and the TO approach in terms of the average downlink sum rate over the testing ground user sets under various values of $W$ and $k$. From Fig. \ref{fig:1_1_delta_p_k_Nued}, we can see that as $k$ rises from 1 to 30 and $W$ varies from 30 to 3000, the performance gap between our mechanism and the TO approach decreases transparently from around $15\times 10^{5}$ bps to almost $0$ bps. Moreover, for a certain value of $W$ or $k$, increasing the value of the other hyper-parameter monotonously improves the proposed mechanism's performance. These observations can be explained as when the proposed UDUA mechanism possesses a larger knowledge database or considers more candidate UAV-BS deployment strategies for a new problem, it will have a higher probability to find the similar ground user sets in the knowledge database and more chances to obtain a proper UDUA solution whose result approaches the optimal value according to \textbf{Proposition 1} and \textbf{Proposition 2}.

An interesting phenomenon in Fig. \ref{fig:1_1_delta_p_k_Nued} is that when $W$ exceeds 500 and $k$ exceeds 5, further augments of $W$ and $k$ will lead to little performance improvement. This is a meaningful conclusion. It not only conﬁrms the practicability of the proposed UDUA mechanism but provides guidance to the hyper-parameter selection as well.

\vspace{-0.0cm}

\subsection{Downlink Sum Rate Performance of the Proposed Mechanism and the Baseline Approaches}
This subsection compares the downlink system throughput achieved by the proposed UDUA mechanism and the baseline approaches. For the proposed mechanism, we set the values of $W$ and $k$ as 500 and 5, respectively, to balance the performance and computational complexity. For the SAUD-GUA approach and the SAUD-KMUA approach, we choose the downlink system throughput as the value of their evaluation functions and set the annealing rate as 0.95. We evaluate the downlink system throughput performance of the five considered approaches under different value combinations of $\mu$ and $\sigma$. Each result is averaged over $N_{\mathrm{Test}}$ testing ground user sets related to a specific network scenario with certain $\mu$ and $\sigma$.

\begin{figure*}
    \centering
    \includegraphics[scale=0.63]{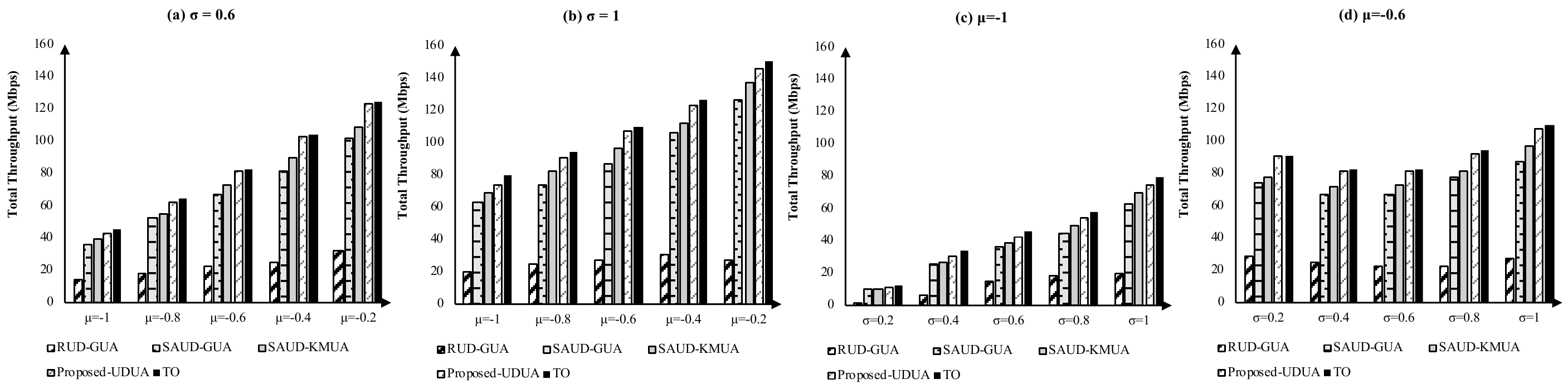}
    \caption{Total transmission throughput comparison among RUD-GUA, SAUD-GUA, SAUD-KMUA, the proposed UDUA, and TO.}
    \label{fig:trans_power}
\end{figure*}

\begin{figure*}
    \centering
    \includegraphics[scale=0.63]{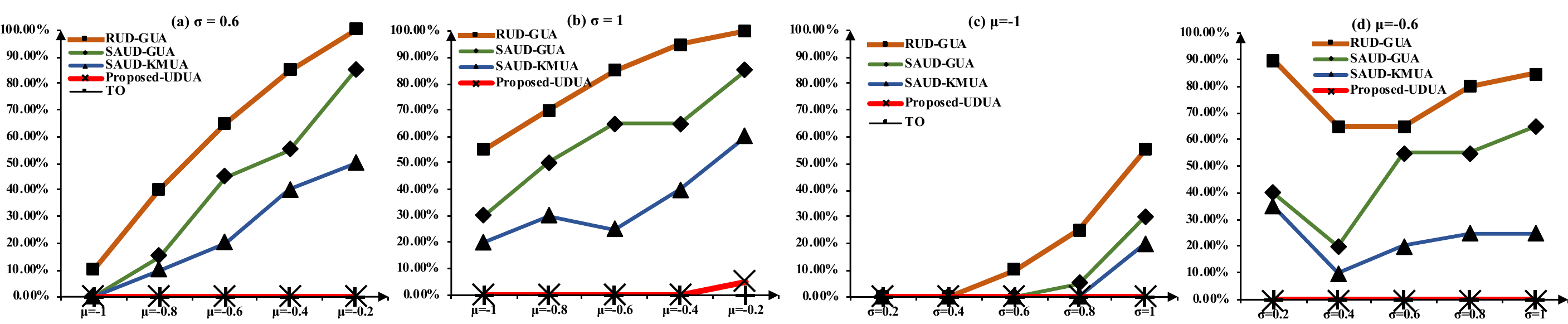}
    \caption{Failure rates comparison among RUD-GUA, SAUD-GUA, SAUD-KMUA, the proposed UDUA, and TO.}
    \label{fig:falure_rates}
\end{figure*}

Fig. \ref{fig:trans_power} (a) and (b) show the average downlink system throughput achieved by the five approaches under different values of $\mu$ when $\sigma=0.6$ or $\sigma=1$. From these figures, we can find that the average downlink system throughput of all the approaches increases as $\mu$ gets large. This is because for a larger $\mu$, the considered region tends to have a larger ground user amount. The RUD-GUA approach causes a very distinct drop in throughput compared with the TO approach (maximum 81.9\% drop when $\sigma=1, \mu=-0.2$). This phenomenon reflects the importance of the UDUA problem addressed in this paper since the ground users might experience very low data rate if the UAV-BSs' locations and associating relationships with the ground users are not assigned properly. Compared with the SAUD-KMUA approach, the SAUD-GUA approach, and the RUD-GUA approach, our UDUA mechanism has the smallest performance gap with the TO under the most experimental scenarios. These numerical results can be explained as follows: first, the proposed mechanism can find the optimal solution of the user association sub-problem to maximize the downlink system throughput for any certain UAV-BS deployment strategy; and second, adopting the optimal UAV-BS deployment strategy of a previous ground user set to a new user set will lead to near optimal performance for the new ground user set if these two sets are similar enough.

Fig. \ref{fig:trans_power} (c) and (d) plot the average downlink system throughput of the five UDUA approaches versus the value of $\sigma$ with $\mu=-1$ and $\mu=-0.6$. The throughput achieved by the five approaches generally ascends as $\sigma$ augments. This can be explained as, besides influencing the non-uniformity of ground user distribution, the increase of $\sigma$ will also raise the user density. With given value of $\mu$, the performance difference between the RUD-GUA approach and the TO approach increases obviously when $\sigma$ gets large. This is because ground users tend to be distributed more non-uniformly in the region $R$ for a bigger $\sigma$, and the positions of UAV-BSs will have a more important effect on the system throughput then. Results in Fig. \ref{fig:trans_power} (c) and (d) also indicate that the proposed UDUA mechanism outperforms the SAUD-KMUA approach by about 10\%-15\% and outperforms the SAUD-GUA approach by about 15\%-20\% with diverse levels of ground user non-uniformity. This can also be owed to new design in the proposed mechanism that the UAV-BS deployment strategy is determined based on the optimal solutions in previous similar UDUA problems and the best ground user association strategy is found with the Kuhn-Munkres algorithm.

\vspace{-0.0cm}

\subsection{Failure Rates of the Proposed Mechanism and the Baseline Approaches}

Fig. \ref{fig:falure_rates} (a) and (b) demonstrate the failure rates of the proposed UDUA mechanism and the baseline approaches under various $\mu$ values with $\sigma=0.6$ and $\sigma=1$. We can see that the five approaches' failure rates increase as $\mu$ grows. These results are consistent with our intuition that a high ground user density will reduce the probability of the fixed number of UAV-BSs to serve all the users successfully and thus lead to a large failure rate.

Besides, the failure rates of approaches with the greedy algorithm based user association (RUD-GUA and SAUD-GUA) ascend more evidently compared with the other approaches. Both of the RUD-GUA and the SAUD-GUA approaches have at least a failure rate of 85\% when $\mu=-0.2$. This is because the greedy algorithm can only find the local optimal user association strategies for each UAV-BS. In RUD-GUA and SAUD-GUA, some ground users may fail to connect to any UAV-BS since the resources are already occupied by other ground users with better channel conditions. Compared with the RUD-GUA, the SAUD-GUA, and the SAUD-KMUA, our mechanism always holds much lower failure rate mainly benefited from the UAV-BS deployment experiences accumulated from well-solved UDUA problems and the optimal user association strategy achieved by the Kuhn-Munkres algorithm. Even when $\sigma$ and $\mu$ have relatively large values ($\sigma=1$, $\mu=-0.2$), our mechanism's failure rate is kept below 5\%.

Fig. \ref{fig:falure_rates} (c) and (d) compare the five approaches' failure rates under different values of $\sigma$ with $\mu=-1$ and $\mu=-0.6$. Similar to the results in (a) and (b), the RUD-GUA and the SAUD-GUA have higher failure rates than the other three approaches and our UDUA mechanism always achieves very low failure rate. An interesting observation in Fig. \ref{fig:falure_rates} (d) is that the failure rates of the RUD-GUA, the SAUD-GUA, and the SAUD-KMUA will first decrease as $\sigma$ augments and then ascend gradually as $\sigma$ continues to increase. This can be explained as follows. On one hand, when $\sigma$ gets large, the ground user will have a higher non-uniformity level and it will be easier for the UAV-BSs to approach the user groups and provide QoS-guaranteed connection services to them. On the other hand, when $\sigma$ exceeds a certain value, the ground user number in region $R$ will become very large and finally dominate the three UDUA approaches' failure rates.

\begin{table*}[]
    \centering
    \scalebox{0.6}{
    \begin{tabular}{c|c|c|c|c|c|c|c|c|c}
\hline
& \multicolumn{3}{c|}{$\mu$=-1} &\multicolumn{3}{c|}{$\mu$=-0.6}&\multicolumn{3}{c}{$\mu$=-0.2}\\
\hline
&$\sigma$=0.2&$\sigma$=0.6&$\sigma$=0.1&$\sigma$=0.2&$\sigma$=0.6&$\sigma$=1&$\sigma$=0.2&$\sigma$=0.6&$\sigma$=1\\
\hline

UDUA-W300-k1&	0.0137466&	0.014254405&	0.016900245&	0.016705055&	0.016148095&	0.017647005&	0.022927455&	0.02042972&	0.02255151\\
UDUA-W300-k10&	0.02289688&	0.03043037&	0.054950465&	0.05456882&	0.04590867&	0.06994845&	0.10083761&	0.089961705&	0.156161265\\
UDUA-W300-k30&	0.04117603&	0.06238793&	0.134665395&	0.13363296&	0.11245862&	0.176088285&	0.255305255&	0.217420615&	0.41938991\\
UDUA-W1000-k1&	0.045704365&	0.04612884&	0.048083285&	0.04742741&	0.04732274&	0.04855999&	0.054944455&	0.049920085&	0.05599631\\
UDUA-W1000-k10&	0.054604205&	0.061711025&	0.08401611&	0.08331422&	0.075963855&	0.099961055&	0.129885865&	0.11558228&	0.175864575\\
UDUA-W1000-k30&	0.07217385&	0.096153265&	0.162838705&	0.171354475&	0.14132068&	0.231497355&	0.30911631&	0.28874234&	0.47164071\\
UDUA-W3000-k1&	0.13345323&	0.13376344&	0.13536682&	0.136121455&	0.135200325&	0.137941375&	0.143128495&	0.13656269&	0.14044655\\
UDUA-W3000-k10&	0.171863755&	0.155750675&	0.17244375&	0.16859234&	0.16186279&	0.18183133&	0.218435375&	0.204592115&	0.245734375\\
UDUA-W3000-k30&	0.159137485&	0.181417615&	0.24988319&	0.248106345&	0.22218461&	0.291448685&	0.389253325&	0.357336505&	0.52918258\\
RUD-GUA	& 0.000133848&	0.000376949&	0.000618136&	0.000754466&	0.000670144&	0.000858879&	0.001078516&	0.001037191&	0.001228497\\
SAUD-GUA&	0.00856747&	0.026085395&	0.051617255&	0.06035117&	0.05228755&	0.080881485&	0.099821695&	0.09867949&	0.153852965\\
SAUD-KMUA&	0.072754425&	0.167257565&	0.71290735&	0.775451995&	0.609347385&	1.682864615&	2.8783834&	3.074763405&	11.93490823\\
TO&	1.9676&	3.8954&	8.29075&	11.1027&	10.8528&	18.93495&	24.27795&	28.1698&	73.50455\\

\hline
    \end{tabular}}
    \caption{Average running time for on-line UDUA problems. The series of UDUA-W-k is the proposed algorithms with different $W$ and $k$. For example, UDUA-W300-k1 represents the proposed UDUA algorithm with $W=300$ and $k=1$. }
    \label{tab:on-line_time_UDUA}
\end{table*}

\begin{table*}[]
    \centering
    \scalebox{0.6}{
    \begin{tabular}{ccccccccccc}
    \hline
    W&	500&	1000&	1500&	2000&	2500&	3000&	3500&	4000&	4500&	5000\\
    \hline
    Storage (kB)&	110&	220&	330&	440&	550&	660&	770&	880&	990&	1100\\
    Off-Line Time (s)  &9805.4083	&16787.6481	&24959.8279	&33441.2987	&41165.9723	&50603.6132	&59041.7728	&69002.8985	&77315.7620	&84990.0406\\
    \hline
    \end{tabular}}
    \caption{The off-line preparing time and storage space of the proposed UDUA algorithm.}
    \label{tab:off-line-time-storage}
\end{table*}

\vspace{-0.0cm}
\subsection{Analyses for Running Time and Storage Space Needed}
We also concern about the running time and storage space needed for the proposed UDUA mechanism. Table II lists the average running time (ART) for on-line UDUA problems of the proposed UDUA mechanism and the baseline approaches under different network scenarios. Specifically, we test the proposed mechanism's ART with various selections of hyper-parameters. From Table \ref{tab:on-line_time_UDUA}, we can find that the RUD-GUA makes the fast decision, which only takes approximately 0.001s since this approach always chooses a random UAV-BS deployment strategy directly and allocates the ground users to UAV-BSs with a low-complexity greedy algorithm. ART of the SAUD-KMUA, the TO, and our mechanism increases when $\mu$ and $\sigma$ gets large. This is because the computational complexity of the proposed bipartite matching theory based solution for the user association sub-problem is positively correlative to the user amount in the considered region. For larger values of $W$ and $k$, our mechanism needs longer running time to search the knowledge database and compare the candidate UAV-BS deployment strategies. However, the on-line running time of the proposed UDUA mechanism is still competitive compared with the SAUD-GUA, the SAUD-KMUA, and the TO, even when $W$ and $k$ have quite large values ($W=3000$, $K=30$).

For each given UDUA problem in the knowledge database, the off-line phase of the proposed mechanism uses the TO approach to find its optimal UAV-BS deployment strategy and then records this UAV-BS deployment strategy as well as the related user distribution matrix. Table \ref{tab:off-line-time-storage} demonstrates the off-line preparation time and storage space needed by the proposed mechanism with different scales of the knowledge database. We can see from Table \ref{tab:off-line-time-storage} that the off-line preparation time and storage space needed are proportional to the value of $W$. Even for a very large $W$ ($W=5000$), the storage space of our mechanism is quite small (less than 1100 KB), and the off-line preparation time needed is acceptable (about 84,990s). Moreover, as analyzed before, although preparing the knowledge database is relatively computing-resource-consuming, we can accomplish this task before the UAV RAN is set.

\vspace{-0.0cm}

\section{Conclusion}
This paper has made an attempt to introduce a simple, fast, and stable machine learning based approach to solve the joint UDUA problems. With the objective of maximizing the downlink sum throughput of all the ground users in a considered region, we formulated the joint UDUA problem as an INLP problem, decoupled it into the user association sub-problem and the UAV-BS deployment sub-problem, and then proposed a centralized UDUA mechanism to solve the two sub-problems respectively. Through extensive simulations with various RAN scenarios, we proved that the proposed UDUA mechanism can achieve near-optimal system performance in terms of average downlink sum transmission rate and failure rate with enormously reduced on-line computing time from hundreds of milliseconds to tens of milliseconds. Furthermore, the optimal hyper-parameter selection of the proposed mechanism has also been analyzed and discussed.

In the future, the channel model with interference will be considered and the dynamic height adjustment of UAV-BSs will be investigated. Also, whether the reinforcement learning technology can be introduced to solve the joint UAV-BS deployment and user association problem is another interesting research direction.


\begin{figure*}
    \centering
    \includegraphics[scale=0.7]{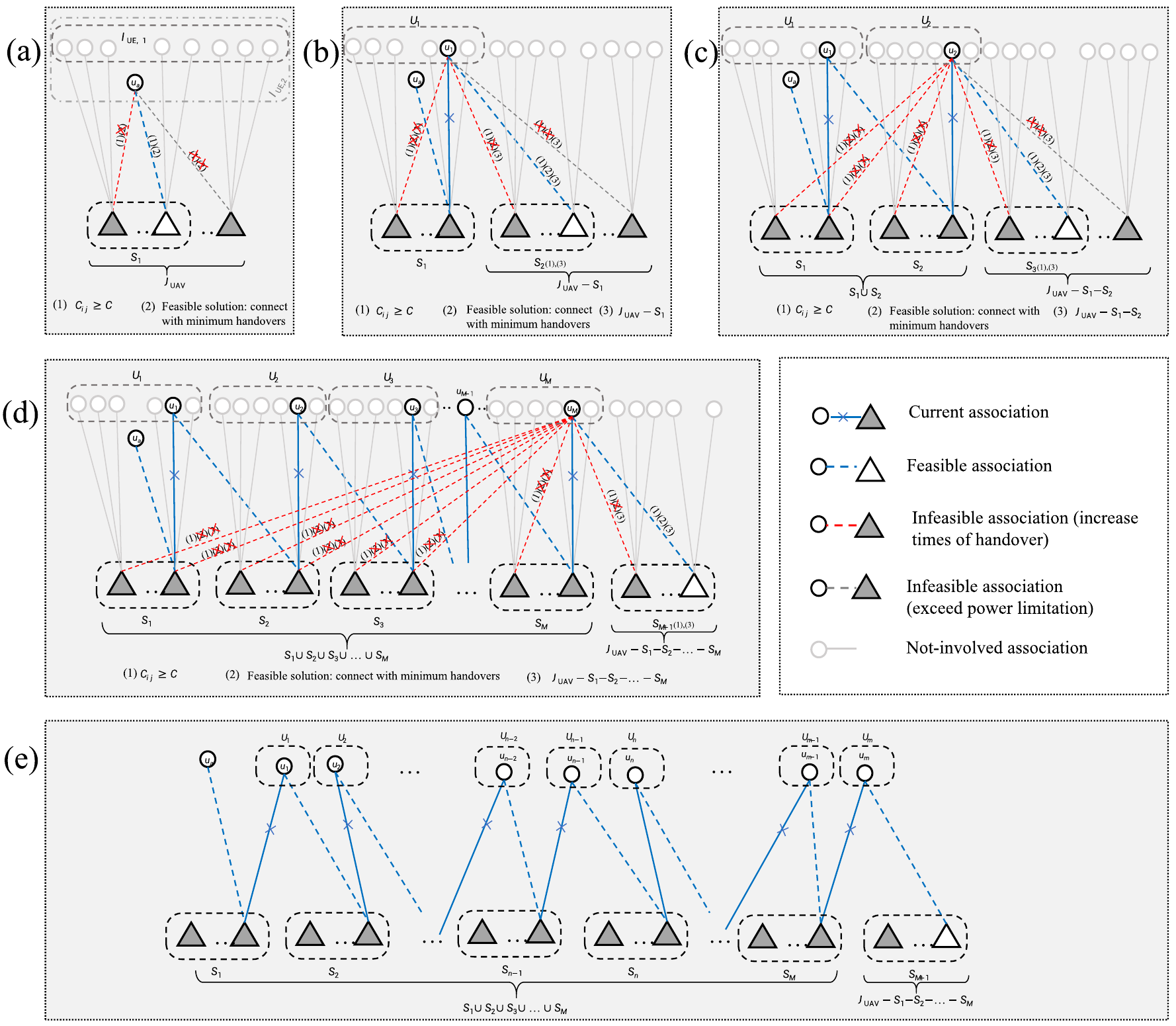}
    \caption{Illustrations of Lemma 1.}
    \label{fig:lemma1}
\end{figure*}

\vspace{-0.0cm}

\appendices
\section{Proof of Lemma 1}
When the UAV-BS deployment strategy $({x}_{1},...,{x}_{J},{y}_{1},...,{y}_{J})$ makes the user association sub-problems related to $I _{\mathrm{UE},1}$ and $I _{\mathrm{UE},2}={I _{\mathrm{UE},1}} \cup \left\{ {{u_a}} \right\}$ have feasible solutions, there are UAV-BSs in set $J _{\mathrm{UAV}}$, whose transmission data rate to ground user $u_a$ will not be less than $C$ if $u_a$ is matched to one of them, and we use set ${S_1} \subseteq {J _{\mathrm{UAV}}}$ to record these UAV-BSs. For an arbitrary feasible user association strategy of $I _{\mathrm{UE},1}$, each ground user in $I _{\mathrm{UE},1}$ will be connected to one UAV-BS and the transmission data rate of the UAV-BS to serve this user should exceed $C$. As demonstrated in Fig. \ref{fig:lemma1}, we will discuss \textbf{Lemma 1} under the following possible conditions:

\begin{enumerate}
    \item As shown in Fig. \ref{fig:lemma1} (a), if there is at least one UAV-BS in set ${S}_1$ possessing available sub-channels in this feasible user association strategy of $I _{\mathrm{UE},1}$, then $u_a$ can be associated to this UAV-BS and the connecting status of no ground user $I _{\mathrm{UE},1}$ will need to be changed. Under this condition, the conclusion of \textbf{Lemma 1} is achieved;
    
    \item Otherwise, if all of the UAV-BSs in set ${S}_1$ are fully occupied by ground users in $I _{\mathrm{UE},1}$, $u_a$ cannot directly be connected to a proper UAV-BS in ${J _{\mathrm{UAV}}}$ with spare sub-channel. We construct the set ${U}_1 \subseteq I _{\mathrm{UE},1}$ to represent the ground users that are associated to the UAV-BSs in ${S}_1$ in the considered feasible user association strategy of $I _{\mathrm{UE},1}$. Since $({x}_{1},...,{x}_{J},{y}_{1},...,{y}_{J})$ makes the user association sub-problem related to $I _{\mathrm{UE},2}={I _{\mathrm{UE},1}} \cup \left\{ {{u_a}} \right\}$ have feasible solutions, there is at least one ground user ${u_1} \in {U_1}$ being connected to a UAV-BS in set ${J _{\mathrm{UAV}}} - {S}_1$ in one feasible user association strategy of $I _{\mathrm{UE},2}$ when $u_a$ is added to one UAV-BS in ${S}_1$. We use the set ${S}_2$ to represent the UAV-BSs, whose transmission data rate to ground user ${u_1}$ should exceed $C$, in set ${J _{\mathrm{UAV}}} - {S_1}$. Obviously, ${S}_2$ is not empty and ${S_2} \cap {S_1} = \phi$, where $\phi$ is an empty set. As demonstrated in Fig. \ref{fig:lemma1} (b), if there is at least one UAV-BS in set ${S}_2$ possessing available sub-channels in this feasible user association strategy of $I _{\mathrm{UE},1}$, then ground user ${u_1}$ can be switched to this UAV-BS and $u_a$ can be associated to the UAV-BS, which previously serves ${u_1}$ in ${S}_1$. Under this condition, the connecting status of one ground user in $I _{\mathrm{UE},1}$ is adjusted and the conclusion of \textbf{Lemma 1} is achieved as the UAV-BS number must be larger than two to construct the sets ${S}_1$ and ${S}_2$;
    
    \item Otherwise, if all of the UAV-BSs in set ${S}_2$ are also fully occupied by ground users in $I _{\mathrm{UE},1}$, we construct the set ${U}_2 \subseteq I _{\mathrm{UE},1}$ to represent the ground users that are associated to the UAV-BSs in ${S}_2$ in the considered feasible user association strategy of $I _{\mathrm{UE},1}$. Because in a feasible user association strategy of $I _{\mathrm{UE},2}$, $u_a$ must be matched with a UAV-BS in ${S_1} \cup {S_2}$ (in ${S_1}$, specifically), there is at least one ground user ${u_2} \in {U_1} \cup {U_2}$ being connected to a UAV-BS in set ${J _{\mathrm{UAV}}} - {S}_1 - {S}_2$ in this user association strategy of $I _{\mathrm{UE},1}$. Obviously, we have ${u_2} \ne {u_1}$. We use the set ${S}_3$ to represent the UAV-BSs, whose transmission data rate to ground user ${u_2}$ exceeds $C$, in set ${J _{\mathrm{UAV}}} - {S_1} - {S_2}$. Also, we have ${S}_3$ is not empty and ${S_3} \cap ({S_1} \cup {S_2}) = \phi$. We set $M=3$. As illustrated in Fig. \ref{fig:lemma1} (c), if there is at least one UAV-BS in set ${S}_3$ possessing available sub-channels in this feasible user association strategy of $I _{\mathrm{UE},1}$, we will go to 5);
    
    \item Otherwise, if all of the UAV-BSs in set ${S}_3$ are fully occupied by ground users in $I _{\mathrm{UE},1}$, we then construct the set ${U}_M$, find the ground user ${u_M} \in {U_1} \cup {U_2} ... \cup {U_M}$ that can be served by a UAV-BS in set ${J _{\mathrm{UAV}}} - {S_1} - {S_2} - ... - {S_M}$ (${u_M} \ne ... \ne {u_2} \ne {u_1}$), construct the set $S_{M+1}$, and judge whether there are UAV-BSs in $S_{M+1}$ possessing available sub-channels in this feasible user association strategy of $I _{\mathrm{UE},1}$, using the similar process in 3). If there is at least one UAV-BS in set ${S}_{M+1}$ possessing available sub-channels, we will go to 5). Otherwise, we will set $M=M+1$ and repeat the above process as depicted in Fig. \ref{fig:lemma1} (d) until there is at least one UAV-BS in set ${S}_{M+1}$ possessing available sub-channels. Since $S_1$, $S_2$, ..., ${S}_{M+1}$ are not empty and at least one UAV-BS in ${J _{\mathrm{UAV}}}$ possessing available sub-channels in this feasible user association strategy of $I _{\mathrm{UE},1}$ ($I _{\mathrm{UE},2}$ will not have feasible user association strategies otherwise), we can finally find the ${S}_{M+1}$ under limited repeats and have $M + 1 \le J$;
    
    \item As shown in Fig. \ref{fig:lemma1} (e), for certain value of $M$, there is a ground user in set ${U_1} \cup {U_2} ... \cup {U_M}$ can be switched to a UAV-BS in ${S}_{M+1}$ possessing available sub-channels. We suppose this ground user belongs to set ${U_m}$ and denote it as ${u_m}$. Obviously, we have $m \le M \le J - 1$. Then, we can associate $u_a$ to a UAV-BS in ${S}_1$ by switching ${u_m}$ to the UAV-BS in ${S}_{M+1}$ possessing available sub-channels, switching ${u_{m-1}}$ to the UAV-BS in ${S}_m$ that previously serves ${u_m}$, ..., switching ${u_1}$ to the UAV-BS in ${S}_2$ that previously serves ${u_2}$, and adding $u_a$ to the UAV-BS in ${S}_1$ that previously serves ${u_1}$. Under this condition, the connecting status of $m$ ground users in $I _{\mathrm{UE},1}$ is adjusted and the conclusion of \textbf{Lemma 1} is achieved as $m \le J - 1$.
\end{enumerate}

Thus, the conclusion of \textbf{Lemma 1} can be achieved under all the conditions. We arrive at \textbf{Lemma 1}.

\vspace{-0.0cm}

\section{Proof of Lemma 2}
For ground user set $I _{\mathrm{UE},1}$ and its optimal UAV-BS deployment strategy $({x}^{*1}_{1},...,{x}^{*1}_{J},{y}^{*1}_{1},...,{y}^{*1}_{J})$, we use ${\Delta ^{*1}} = (\delta _{ij}^{*1},\mathrm{UE}_i \in {I _{\mathrm{UE},1}},\mathrm{UAV}_j \in {J _{\mathrm{UAV}}})$ to represent the optimal solution of the related user association sub-problem. Obviously, ${\Delta ^{*1}}$ is a feasible user association strategy of $I _{\mathrm{UE},1}$.

Without loss of generality, we denote the $m$ new ground users in $I _{\mathrm{UE},2}$ as ${u_{\mathrm{new},1}}$, ${u_{\mathrm{new},2}}$, ..., and ${u_{\mathrm{new},m}}$. Since $({x}^{*1}_{1},...,{x}^{*1}_{J},{y}^{*1}_{1},...,{y}^{*1}_{J})$ makes the user association sub-problem related to $I _{\mathrm{UE},2}$ have feasible solutions, this UAV-BS deployment strategy will also make the user association sub-problems related to ${I _{\mathrm{UE},1}} \cup \{ {u_{\mathrm{new},1}}\}$, ${I _{\mathrm{UE},1}} \cup \{ {u_{\mathrm{new},1}},{u_{\mathrm{new},2}}\}$, and ${I _{\mathrm{UE},1}} \cup \{ {u_{\mathrm{new},1}},{u_{\mathrm{new},2}},...,{u_{\mathrm{new},(m - 1)}}\}$ have feasible solutions. According to \textbf{Lemma 1}, we can connect ${u_{\mathrm{new},1}}$ to a proper UAV-BS and find a feasible user association strategy of ${I _{\mathrm{UE},1}} \cup \{ {u_{\mathrm{new},1}}\}$, ${\Delta _{{I _{\mathrm{UE},1}} \cup \{ {u_{\mathrm{new},1}}\} }}$, from ${\Delta ^{*1}}$ by adjusting the connecting statuses of up to $J-1$ previous ground users. Because the downlink throughput changes caused by serving ${u_{\mathrm{new},1}}$ or changing the associated UAV-BS of a previous ground user is ${\epsilon _{\mathrm{min} }}$ or $({\epsilon_{\mathrm{max} }} - {\epsilon _{\mathrm{min} }})$, respectively, we have the following inequality:
\begin{equation}
\begin{array}{l}
C({\Delta _{{I _{\mathrm{UE},1}} \cup \{ {u_{\mathrm{new},1}}\} }}) \ge C({\Delta ^{*1}}) + {\epsilon _{\mathrm{min} }} -\\
(J - 1)  ({\epsilon _{\mathrm{max} }} - {\epsilon _{\mathrm{min} }})\\
 = {f_{{I _{\mathrm{UE},1}}}}(x_1^{*1},\;...,\;x_J^{*1},\;y_1^{*1},\;...,\;y_J^{*1}) + {\epsilon _{\mathrm{min} }} - \\
(J - 1)  ({\epsilon _{\mathrm{max} }} - {\epsilon _{\mathrm{min} }}),
\end{array}
\label{eq:appendixB_29}
\end{equation}

\noindent
where $C({\Delta _{{I _{\mathrm{UE},1}} \cup \{ {u_{\mathrm{new},1}}\} }})$ and $C({\Delta ^{*1}})$ are the values of downlink throughput related to ${\Delta _{{I _{\mathrm{UE},1}} \cup \{ {u_{\mathrm{new},1}}\} }}$ and ${\Delta ^{*1}}$, respectively. Similarly, we can also prove the following inequalities:
\begin{equation}
\begin{array}{l}
C({\Delta _{{I _{\mathrm{UE},1}} \cup \{ {u_{\mathrm{new},1}},{u_{\mathrm{new},2}}\} }}) \ge C({\Delta _{{I _{\mathrm{UE},1}} \cup \{ {u_{\mathrm{new},1}}\} }}) + \\
{\epsilon _{\min }} - (J - 1)  ({\epsilon _{\max }} - {\epsilon _{\min }}) \; ...\\
C({\Delta _{{I _{\mathrm{UE},2}}}}) \ge C({\Delta _{{I _{\mathrm{UE},1}} \cup \{ {u_{\mathrm{new},1}},{u_{\mathrm{new},2}},...,{u_{\mathrm{new},(m - 1)}}\} }}) + \\
{\epsilon _{\min }} - (J - 1)  ({\epsilon _{\max }} - {\epsilon _{\min }}),
\end{array}
\label{eq:appendixB_30}
\end{equation}

\noindent
where ${\Delta _{{I _{\mathrm{UE},1}} \cup \{ {u_{\mathrm{new},1}},{u_{\mathrm{new},2}}\} }}$, ..., ${\Delta _{{I _{\mathrm{UE},1}} \cup \{ {u_{\mathrm{new},1}},{u_{\mathrm{new},2}},...,{u_{\mathrm{new},(m - 1)}}\} }}$, and ${\Delta _{{I _{\mathrm{UE},2}}}}$ are the feasible user association strategies of ${I _{\mathrm{UE},1}} \cup \{ {u_{\mathrm{new},1}},{u_{\mathrm{new},2}}\}$, ..., ${I _{\mathrm{UE},1}} \cup \{ {u_{\mathrm{new},1}},{u_{\mathrm{new},2}},...,{u_{\mathrm{new},(m - 1)}}\}$, and $I _{\mathrm{UE},2}$, respectively.
Furthermore, we have ${f_{{I _{\mathrm{UE},2}}}}(x_1^{*1},\;...,\;x_J^{*1},\;y_1^{*1},\;...,\;y_J^{*1}) \ge C({\Delta _{{I _{\mathrm{UE},2}}}})$ due to the fact that ${f_{{I _{\mathrm{UE},2}}}}(x_1^{*1},\;...,\;x_J^{*1},\;y_1^{*1},\;...,\;y_J^{*1})$ is the optimal value of user association sub-problem related to $I _{\mathrm{UE},2}$ when the UAV-BS deployment strategy is $({x}^{*1}_{1},...,{x}^{*1}_{J},{y}^{*1}_{1},...,{y}^{*1}_{J})$. Thus, we can get (\ref{eq:lemma2-21}) through (\ref{eq:appendixB_29}) and (\ref{eq:appendixB_30}). \textbf{Lemma 2} is proved.

\vspace{-0.0cm}

\section{Proof of Proposition 1}
When $I _{\mathrm{UE},2}$ is obtained by adding $m$ new ground users into $I _{\mathrm{UE},1}$, we denote $I _{\mathrm{UE},2}$ as ${I _{\mathrm{UE},1}} \cup \{ {u_{\mathrm{new},1}},{u_{\mathrm{new},2}},...,{u_{\mathrm{new},m}}\}$ without loss of generality. Since the transmission data rate of an arbitrary UAV-BS in set ${J _{\mathrm{UAV}}}$ to serve a ground user is not larger than ${\epsilon _{\mathrm{max} }}$, we have the following inequality:
\begin{equation}
\begin{array}{l}
{f_{{I _{\mathrm{UE},2}}}}(x_1^{*2},...,x_J^{*2},y_1^{*2},...,y_J^{*2}) \le \\
m  {\epsilon _{\max }} + {f_{{I _{\mathrm{UE},1}}}}(x_1^{*2},...,x_J^{*2},y_1^{*2},...,y_J^{*2}),
\end{array}
\label{eq:appendixC_31}
\end{equation}

\noindent
where ${f_{{I _{\mathrm{UE},1}}}}(x_1^{*2},...,x_J^{*2},y_1^{*2},...,y_J^{*2})$ is the optimal value of user association sub-problem related to $I _{\mathrm{UE},1}$ when the UAV-BS deployment strategy is $x_1^{*2},...,x_J^{*2},y_1^{*2},...,y_J^{*2}$. Since $({x}^{*1}_{1},...,{x}^{*1}_{J},{y}^{*1}_{1},...,{y}^{*1}_{J})$ is the optimal UAV-BS deployment strategy for $I _{\mathrm{UE},1}$, the following inequality can be achieved:
\begin{equation}
\begin{array}{l}
{f_{{I _{\mathrm{UE},1}}}}(x_1^{*1},\;...,\;x_J^{*1},\;y_1^{*1},\;...,\;y_J^{*1}) \ge \\
{f_{{I _{\mathrm{UE},1}}}}({x}^{*2}_{1},...,{x}^{*2}_{J},{y}^{*2}_{1},...,{y}^{*2}_{J}).
\end{array}
\label{eq:appendixC_32}
\end{equation}

According to \textbf{Lemma 2}, we have:
\begin{equation}
\begin{array}{l}
{f_{{I _{\mathrm{UE},2}}}}(x_1^{*1},\;...,\;x_J^{*1},\;y_1^{*1},\;...,\;y_J^{*1}) \ge \\
{f_{{I _{\mathrm{UE},1}}}}(x_1^{*1},\;...,\;x_J^{*1},\;y_1^{*1},\;...,\;y_J^{*1}) + \\
m  [\epsilon _{\mathrm{min}} - (J - 1)  ({\epsilon _{\mathrm{max} }} - {\epsilon _{\mathrm{min} }})].
\end{array}
\label{eq:appendixC_33}
\end{equation}

By jointly considering (\ref{eq:appendixC_31}), (\ref{eq:appendixC_32}), and (\ref{eq:appendixC_33}), we can get (\ref{eq:proposition1-22}) immediately.

When $I _{\mathrm{UE},2}$ is obtained by removing $m$ ground users off $I _{\mathrm{UE},1}$, we denote $I _{\mathrm{UE},1}$ as ${I _{\mathrm{UE},2}} \cup \{ {u_{\mathrm{new},1}},{u_{\mathrm{new},2}},...,{u_{\mathrm{new},m}}\}$ without loss of generality. According to \textbf{Lemma 2}, we have:
\begin{equation}
\begin{array}{*{20}{l}}
{{f_{{I _{\mathrm{UE},1}}}}(x_1^{*2},\;...,\;x_J^{*2},\;y_1^{*2},\;...,\;y_J^{*2}) \ge }\\
{{f_{{I _{\mathrm{UE},2}}}}(x_1^{*2},\;...,\;x_J^{*2},\;y_1^{*2},\;...,\;y_J^{*2}) + }\\
{m  [{\epsilon _{\mathrm{min}}} - (J - 1)  ({\epsilon _{\mathrm{max}}} - {\epsilon _{\mathrm{min}}})]}.
\end{array}
\label{eq:appendixC_34}
\end{equation}

Furthermore, since the transmission data rate of an arbitrary UAV-BS in set ${J _{\mathrm{UAV}}}$ to serve a ground user is not less than ${\epsilon _{\mathrm{min} }}$, we have the following inequality:
\begin{equation}
\begin{array}{*{20}{l}}
{{f_{{I _{\mathrm{UE},1}}}}(x_1^{*1},\;...,\;x_J^{*1},\;y_1^{*1},\;...,\;y_J^{*1}) \le }\\
{m  {\epsilon_{\max}} + {f_{{I _{\mathrm{UE},2}}}}(x_1^{*1},\;...,\;x_J^{*1},\;y_1^{*1},\;...,\;y_J^{*1})}.
\end{array}
\label{eq:appendixC_35}
\end{equation}

Using the inequality in (\ref{eq:appendixC_32}) again and combining (\ref{eq:appendixC_34}) with (\ref{eq:appendixC_35}), we can get (\ref{eq:proposition1-22}) immediately.

So when $I _{\mathrm{UE},2}$ is acquired by adding $m$ ground users into or removing $m$ ground users off $I _{\mathrm{UE},1}$, (\ref{eq:proposition1-22}) can be satisfied. We arrive at \textbf{Proposition 1}.

\vspace{-0.0cm}

\section{Proof of Lemma 3}
When $n$ ground users in $I _{\mathrm{UE},1}$ change their position grids in region $R$ and generate $I _{\mathrm{UE},2}$, we denote $I _{\mathrm{UE},1}$ as ${I _{\mathrm{UE},\mathrm{stable}}} \cup \{ {u_{\mathrm{move},1}},{u_{\mathrm{move},2}},...,{u_{\mathrm{move},n}}\}$. ${I _{\mathrm{UE},\mathrm{stable}}}$ is the set of ground users in $I _{\mathrm{UE},1}$ remaining stable and $\{ {u_{\mathrm{move},1}},{u_{\mathrm{move},2}},...,{u_{\mathrm{move},n}}\}$ is the set of ground users who will move inside $R$. After the ground users in $\{ {u_{\mathrm{move},1}},{u_{\mathrm{move},2}},...,{u_{\mathrm{move},n}}\}$ have been allocated at their new positions, we denote $I _{\mathrm{UE},2}$ as ${I _{\mathrm{UE},\mathrm{stable}}} \cup \{ u_{\mathrm{move},1}^{'},u_{\mathrm{move},2}^{'},...,u_{\mathrm{move},n}^{'}\}$. Since the transmission data rate of an arbitrary UAV-BS in set ${J _{\mathrm{UAV}}}$ to serve a ground user can not exceed ${\epsilon _{\mathrm{max} }}$, we have the following inequality:
\begin{equation}
\begin{array}{l}
{f_{{I _{\mathrm{UE},1}}}}(x_1^{*1},\;...,\;x_J^{*1},\;y_1^{*1},\;...,\;y_J^{*1}) \le \\
n  {\epsilon _{\mathrm{max}}} + {f_{{I _{\mathrm{UE},\mathrm{stable}}}}}(x_1^{*1},\;...,\;x_J^{*1},\;y_1^{*1},\;...,\;y_J^{*1}),
\end{array}
\label{eq:appendixD_36}
\end{equation}

\noindent
where ${f_{{I _{\mathrm{UE},\mathrm{stable}}}}}(x_1^{*1},\;...,\;x_J^{*1},\;y_1^{*1},\;...,\;y_J^{*1})$ is the optimal value of user association sub-problem related to ${I _{\mathrm{UE},\mathrm{stable}}}$ when the UAV-BS deployment strategy is $(x_1^{*1},\;...,\;x_J^{*1},\;y_1^{*1},\;...,\;y_J^{*1})$.

Since $I _{\mathrm{UE},2}$ can be regarded as the ground user set obtained by adding the $n$ ground users in $\{ u_{\mathrm{move},1}^{'},u_{\mathrm{move},2}^{'},...,u_{\mathrm{move},n}^{'}\}$, we achieve the following inequality based on \textbf{Lemma 2}:
\begin{equation}
\begin{array}{l}
{f_{{I _{\mathrm{UE},2}}}}(x_1^{*1},\;...,\;x_J^{*1},\;y_1^{*1},\;...,\;y_J^{*1}) \ge \\
{f_{{I _{\mathrm{UE},\mathrm{stable}}}}}(x_1^{*1},\;...,\;x_J^{*1},\;y_1^{*1},\;...,\;y_J^{*1}) + \\
n  [{\epsilon _{\mathrm{min}}} - (J - 1)  ({\epsilon _{\mathrm{max}}} - {\epsilon _{\mathrm{min}}})].
\end{array}
\label{eq:appendixD_37}
\end{equation}

Combining (\ref{eq:appendixD_36}) with (\ref{eq:appendixD_37}), we will get (\ref{eq:lemma3-23}). Thus, \textbf{Lemma 3} is proved.

\section{Proof of Proposition 2}

When the UAV-BS deployment is fixed to $({x}^{*2}_{1},...,{x}^{*2}_{J},{y}^{*2}_{1},...,{y}^{*2}_{J})$, by following Lemma 3, we can have this inequality,
\begin{equation}
\begin{array}{*{20}{l}}
{{f_{{I _{\mathrm{UE},1}}}}({x}^{*2}_{1},...,{x}^{*2}_{J},{y}^{*2}_{1},...,{y}^{*2}_{J}) \ge }\\
{{f_{{I _{\mathrm{UE},2}}}}({x}^{*2}_{1},...,{x}^{*2}_{J},{y}^{*2}_{1},...,{y}^{*2}_{J}) - }\\
{n  J  ({\epsilon _{\mathrm{max}}} - {\epsilon _{\mathrm{min}}})}.
\end{array}
\label{eq:appendixE_38}
\end{equation}

That is because transforming $I _{\mathrm{UE},1}$ to $I _{\mathrm{UE},2}$ is symmetrical to transforming $I _{\mathrm{UE},2}$ to $I _{\mathrm{UE},1}$. The number of moved UEs is same as $n$. 

When the UAV-BSs' locations are changed, ${{f_{{I_{\mathrm{UE},1}}}}({x}^{*2}_{1},...,{x}^{*2}_{J},{y}^{*2}_{1},...,{y}^{*2}_{J})  }$ cannot be less than the optimum solution ${{f_{{I _{\mathrm{UE},1}}}}(x_1^{*1},\;...,\;x_J^{*1},\;y_1^{*1},\;...,\;y_J^{*1})  }$:
\begin{equation}
\begin{array}{*{20}{l}}
{{f_{{I _{\mathrm{UE},1}}}}(x_1^{*1},\;...,\;x_J^{*1},\;y_1^{*1},\;...,\;y_J^{*1}) \ge }\\
{{f_{{I _{\mathrm{UE},1}}}}({x}^{*2}_{1},...,{x}^{*2}_{J},{y}^{*2}_{1},...,{y}^{*2}_{J})  }.
\end{array}
\label{eq:appendixE_39}
\end{equation}

According to (\ref{eq:lemma3-23}), (\ref{eq:appendixE_38}), and (\ref{eq:appendixE_39}), the following inequality is derived:
\begin{equation}
\begin{array}{*{20}{l}}
{{f_{{I _{\mathrm{UE},2}}}}(x_1^{*1},\;...,\;x_J^{*1},\;y_1^{*1},\;...,\;y_J^{*1}) \ge }\\
{{f_{{I _{\mathrm{UE},2}}}}(x_1^{*2},\;...,\;x_J^{*2},\;y_1^{*2},\;...,\;y_J^{*2}) - }
{2  n  J  ({\epsilon _{\mathrm{max}}} - {\epsilon _{\mathrm{min}}})}.
\end{array}
\end{equation}

We arrive at \textbf{Proposition 2}.


\ifCLASSOPTIONcaptionsoff
  \newpage
\fi

\bibliographystyle{myIEEEtran}
\end{document}